\documentclass[sigconf,authorversion,natbib=true,anonymous=false,hyperref={breaklinks=True}]{acmart}

\usepackage[normalem]{ulem}
\usepackage{booktabs} 
\usepackage{color}
\usepackage{balance}

\copyrightyear{2022} 
\acmYear{2022} 
\setcopyright{acmcopyright}\acmConference[SIGIR '22]{Proceedings of the 45th International ACM SIGIR Conference on Research and Development in Information Retrieval}{July 11--15, 2022}{Madrid, Spain}
\acmBooktitle{Proceedings of the 45th International ACM SIGIR Conference on Research and Development in Information Retrieval (SIGIR '22), July 11--15, 2022, Madrid, Spain}
\acmPrice{15.00}
\acmDOI{10.1145/3477495.3531739}
\acmISBN{978-1-4503-8732-3/22/07}

\begin{document}

\fancyhead{}
\title{Would You Ask it that Way? Measuring and Improving Question Naturalness for Knowledge Graph Question Answering}

\author{Trond Linjordet}
\affiliation{
  \institution{University of Stavanger}
  \city{Stavanger}
  \country{Norway}
}
\email{trond.linjordet@uis.no}

\author{Krisztian Balog}
\affiliation{
  \institution{University of Stavanger}
  \city{Stavanger}
  \country{Norway}
}
\email{krisztian.balog@uis.no}

\begin{abstract}
Knowledge graph question answering (KGQA) facilitates information access by leveraging structured data without requiring formal query language expertise from the user.  Instead, users can express their information needs by simply asking their questions in natural language (NL).  Datasets used to train KGQA models that would provide such a service are expensive to construct, both in terms of expert and crowdsourced labor.  
Typically, crowdsourced labor is used to improve template-based pseudo-natural questions generated from formal queries.
However, the resulting datasets often fall short of representing genuinely natural and fluent language. 
In the present work, we investigate ways to characterize and remedy these shortcomings.  We create the IQN-KGQA test collection by sampling questions from existing KGQA datasets and evaluating them with regards to five different aspects of naturalness.  Then, the questions are rewritten to improve their fluency.  Finally, the performance of existing KGQA models is compared on the original and rewritten versions of the NL questions.  
We find that some KGQA systems fare worse when presented with more realistic formulations of NL questions.  
The IQN-KGQA test collection is a resource to help evaluate KGQA systems in a more realistic setting.  
The construction of this test collection also sheds light on the challenges of constructing large-scale KGQA datasets with genuinely NL questions. 
\end{abstract}




\maketitle
	
\begin{table}[t]
\vspace*{1.5\baselineskip}
\caption{Example questions, each rewritten by crowd workers as a more natural way to express the original question.} 
\label{tab:examples}
\centering
\begin{tabular}{p{3.9cm}p{3.9cm}} 
    \toprule
    \textbf{Original question} & \textbf{Rewritten question} \\
    (DBNQA~\cite{Hartmann:2018:WEBBR-ws}) & (IQN-KGQA [this paper]) \\
    \midrule
    List the territory of romanian war of independence ? & What territory was involved in the Romanian War of Independence? \\
    Name the nearest city to la laguna lake ? & What is the nearest city to La Laguna Lake? \\
    What is the government type of wallis and futuna ? & What type of government does Wallis and Futuna have? \\
    What is the origin of faberrebe? & What is the origin of the faberrebe grape? \\
    What is the total number of writers whose singles are recorded in ferndale? & How many writers had singles recorded in Ferndale? \\
    \bottomrule
\end{tabular}   
\end{table}

\section{Introduction}

Knowledge Graph Question Answering (KGQA) is an approach to answering users' questions that both harnesses structured data in the form of knowledge graphs (KGs) and also allows the user to articulate their information need in natural language (NL). 
Training machine learning models for KGQA requires large-scale datasets specific to the KGQA task. Most commonly, such datasets consist of instances that each comprises a formal query (also known as logic form) and a corresponding NL question~\citep{Lan:2021:IJCAI}. 

In order to construct large KGQA datasets, the work is typically divided into expert and non-expert subtasks which are then assigned to different people. 
This makes sense economically, but the resulting dataset may have qualitative shortcomings as a result. 
The formal query is typically constructed by experts or generated synthetically, while the NL questions are typically added by crowdsourced labor tasked with paraphrasing some generated pseudo-natural form of the corresponding formal query~\citep{Wang:2015:ACL-IJNLP}. 
The NL question is thus typically not formulated by the same person who devised the formal query. 
Critically, this decouples the intent of the formal query from the NL question meant to express that intent.
In addition, in large-scale dataset construction, the data is often back-generated from formal queries, that is, the formal query is generated based on available data, and the corresponding NL question is created afterwards.
An individual working with a KG for practical reasons would first develop an information need, which may or may not be first expressed as an NL question, and only then construct a formal query to represent that information need. 
The crowd worker is also not guaranteed to be completely fluent in the specific language that is used in the dataset being constructed. 
Furthermore, even so-called open-domain KGQA typically consists of questions in a variety of specific domains. 
If the crowd worker is unfamiliar with this domain, they may not be able to apply the appropriate wording for the underlying domain and categories. 
We therefore hypothesize that this approach to KGQA dataset construction does not ensure genuinely natural NL questions. 

In Table~\ref{tab:examples} we have listed three example questions sampled from existing KGQA datasets, each in their original form and in a rewritten form, generated by additional rounds of crowdsourced paraphrasing and quality control. These are all examples where a KGQA model trained on the original dataset performed perfectly on the original question, but completely failed on the rewritten question. This illustrates that some KGQA systems trained on less natural NL questions are not able to address a more naturally phrased version of the same question.

From a machine learning perspective, it is unsurprising that test data from a different distribution than training data may be challenging. However, as the rewritten questions in Table~\ref{tab:examples} illustrate, the KGQA models are failing on more naturally articulated questions. This calls into question whether KGQA models are really learning to perform their nominal task.  
We investigate how NL questions in KGQA datasets can be considered unnatural, and develop a coding scheme for dimensions of unnaturalness. 
We determine five dimensions of unnaturalness in NL questions: grammar, form, meaning, answerability, and factuality. 

Next, we use our coding scheme in a crowdsourcing context to characterize original NL questions and collect rewritten forms of these NL questions, which are included in our test collection, IQN-KGQA. We sample 250 NL questions from each of three benchmark KGQA datasets: DBNQA~\citep{Hartmann:2018:WEBBR-ws}, LC-QuAD v2.0~\citep{Dubey:2019:ISWC}, and GrailQA~\citep{Gu:2021:WWW}. 

To develop truly effective KGQA systems requires an appreciation of how well these systems fare against realistic questions formulated in genuinely natural language.
We apply KGQA models to the original and rewritten questions and see how improved naturalness challenges existing systems. We find that performance drops up to 78\% when KGQA models are challenged with the set of rewritten questions. 

The novel contributions of this work include:
\begin{itemize}
	\item A novel coding scheme to characterize to what extent nominal natural language questions in KGQA datasets actually constitute natural language.
	\item Experimental designs for measuring question naturalness (to flag questions that are unnatural) and improving question fluency and composition using crowdsourcing.
	\item A novel test collection, IQN-KGQA, consisting of 3x250 questions sampled from 3 prominent KGQA datasets, made publicly available at \url{https://github.com/iai-group/IQN-KGQA}. The sampled questions are rated on naturalness along 5 dimensions by at least 3 crowd workers each, and rewritten for greater naturalness where possible.
	\item A comparison of existing KGQA models on original and rewritten questions.
\end{itemize}
\begin{table*}[t]
\caption{Overview of KGQA datasets. Those marked with $^\dag$ are considered in our study.}
\label{tab:datasets}
\centering
\begin{tabular}{l lrc} 
    \toprule
    \textbf{Dataset} & \textbf{KG} & \textbf{Size} & \textbf{Crowdsourcing} \\ 
    & & & \\ 
    \midrule
    Free917~\citep{Cai:2013:ACL} & Freebase & 917 & Unclear \\
    WebQuestions~\citep{Berant:2013:EMNLP} & Freebase & 5,810 & Yes \\
    SimpleQuestions~\citep{Bordes:2015:arXiv} & Freebase & 108,442 & Unclear \\
    ComplexQuestions~\citep{Bao:2016:COLING} & Freebase & 2,100 & No \\
    GraphQuestions~\citep{Su:2016:EMNLP} & Freebase & 5,166 &  Yes \\
    WebQuestionsSP~\citep{Yih:2016:ACL} & Freebase & 4,737 & No \\
    ComplexWebQuestions~\citep{Talmor:2018:NAACL} & Freebase & 34,689 & Yes \\ 
    QALD series (1--9)~\citep{Lopez:2013:WS, Usbeck:2017:ESWC} & DBpedia & $\sim$50-500 each & No \\ 
    LC-QuAD v1.0~\citep{Trivedi:2017:ISWC} & DBpedia & 5,000 & No \\
    DBNQA~\citep{Hartmann:2018:WEBBR-ws}$^\dag$ & DBpedia & 894,499 & No \\
    LC-QuAD v2.0~\citep{Dubey:2019:ISWC}$^\dag$ & DBpedia, Wikidata & 30,000 & Yes \\
    CFQ~\citep{Keysers:2020:ICLR} & Freebase & 239,357 & No \\
    GrailQA~\citep{Gu:2021:WWW}$^\dag$ & Freebase & 64,331 & Yes \\
    KQA Pro~\citep{Shi:2020:arXiv} & Wikidata & 117,970 & Yes \\ 
    \bottomrule
\end{tabular}    
\end{table*}

\section{Related work}
\label{sec:related}

The field of KGQA is in many ways defined by the datasets used to train and test systems. In the following, we describe some salient milestone KGQA datasets and their manner of construction. Previous work~\citep{Chakraborty:2020:WIREsDMKD, Lan:2021:IJCAI, Roy:2021:SLcICRS, Wu:2019:CCKS} has surveyed the field of KGQA, which we draw on in our present summary. The KGQA datasets are grounded in one or more of the three most common open-domain knowledge graphs (KGs): Freebase, DBpedia, and Wikidata. We note that the overall trend in KGQA dataset construction has been towards more complex formal queries as well as larger datasets. For each dataset, we defer to the respective papers' stance as to whether the dataset should be considered to contain complex formal queries.

\citet{Cai:2013:ACL} create the dataset Free917 by asking two native English speakers to ask questions in multiple domains, and then annotating these questions with formal queries. 

\citet{Berant:2013:EMNLP} construct the dataset WebQuestions, consisting of 5810 instances with only NL questions and answers, but no formal queries. The dataset is constructed by generating single-entity questions with the Google Suggest API, and then crowdsourcing answers based only on the Freebase page of the entity in a given NL question. Question-answer pairs are kept as instances when at least two crowd workers agree on an answer.

\citet{Bordes:2015:arXiv} create the large dataset SimpleQuestions, consisting only of NL questions that can be answered by a single fact (SPO-triple) in the KG, and the corresponding fact. The dataset is constructed by shortlisting a set of facts, and then having English-speaking annotators generate NL questions mentioning the subject and predicate of the fact, such that the answer would be the object. 

\citet{Bao:2016:COLING} construct the dataset ComplexQuestions consisting of question-answer pairs by mining a search query log for search queries with overlapping terms as in WebQuestions and SimpleQuestions, and then categorize the search queries according to some rules to identify multi-constraint questions. The questions are manually annotated with answers. Additional question-answer pairs are taken directly from pre-existing datasets.  

\citet{Su:2016:EMNLP} construct the dataset GraphQuestions---where each instance includes NL question, formal query, and ground truth answer---by first generating query graphs, and then converting these to NL questions via crowdsourcing. The ground truth answer is retrieved by converting the query graph to a formal query and executing it. This approach to crowdsourcing for KGQA datasets has been referred to as the Overnight method~\citep{Su:2016:EMNLP}. 

\citet{Yih:2016:ACL} construct the dataset WebQuestionsSP by having experts annotate instances in WebQuestions~\citep{Berant:2013:EMNLP} with SPARQL queries where feasible. 

\citet{Talmor:2018:NAACL} construct the dataset ComplexWebQuestions by programmatically generating more complex formal queries from WebQuestionsSP, then generating pseudo-NL questions for crowd workers to improve into NL questions. 

The QALD series (1--9)~\citep{Lopez:2013:WS} consists of small datasets of questions generated by students and formal queries hand-crafted by experts. 

\citet{Trivedi:2017:ISWC} construct the dataset LC-QuAD v1.0, which consists of NL questions and formal queries. First query graph templates are combined with whitelisted (non-metadata) entities and predicates to instantiate specific formal queries, then pseudo-NL questions are generated from the formal queries, which are corrected or paraphrased in two rounds with independent annotators.

\citet{Hartmann:2018:WEBBR-ws} construct the dataset DBNQA, consisting of NL questions and formal queries, from the LC-QuAD v1.0~\citep{Trivedi:2017:ISWC} and QALD-7-train~\citep{Usbeck:2017:ESWC} datasets. The extant datasets are taken as the basis to extract templates for both formal queries and NL questions, and those templates are then instantiated with different entity and predicate bindings. 
DBNQA*~\citep{Linjordet:2020:ICTIR} partitions DBNQA~\citep{Hartmann:2018:WEBBR-ws} into training, validation, and test splits based on the underlying templates, avoiding leakage of information between training and test splits. The instances are identical to DBNQA, and so we use DBNQA* in our experiments.

\citet{Dubey:2019:ISWC} construct the dataset LC-QuAD v2.0, extending the workflow established by \citet{Trivedi:2017:ISWC} by crowdsourcing the paraphrasing of generated pseudo-NL questions into improved NL questions. This also includes several rounds of crowd workers generating further paraphrasing of NL questions and performing quality control on others' annotations. 

\citet{Keysers:2020:ICLR} construct the dataset CFQ, with instances comprising formal queries and and NL questions, in a completely rules-based manner. 

\citet{Gu:2021:WWW} construct the dataset GrailQA following the Over\-night~\citep{Su:2016:EMNLP} approach of generating formal queries and pseudo-NL questions, and then using crowdsourcing to paraphrase pseudo-NL questions into NL questions, and finally using crowd workers to cross-validate the paraphrases of their colleagues. 

\citet{Shi:2020:arXiv} construct the dataset KQA Pro in a similar manner as \citet{Gu:2021:WWW}, including the use of crowdsourced labor for paraphrasing pseudo-NL questions and cross-validation.  

From these examples of KGQA dataset construction, summarized in Table~\ref{tab:datasets}, we see that crowdsourcing is commonly used with the intent of paraphrasing pseudo-NL questions into more genuine NL questions. 
\section{Preliminary Analysis}
\label{sec:prelim}

Larger KGQA datasets typically rely on crowdsourcing for generating NL questions from synthetically generated formal queries. We hypothesize that scaling up a KGQA dataset by relying heavily on these two distinct modes comes at the expense of NL question quality, and that the original NL questions may not always be genuinely natural NL questions. To investigate unnaturalness in the NL questions of existing KGQA datasets, we begin by testing that hypothesis on a small sample of instances using expert annotators.

\begin{table*}[t!]
\caption{Dimensions of question unnaturalness}
\label{tab:coding}
\centering
\begin{tabular}{llp{10cm}} 
    \toprule
    \textbf{Dimension} & \textbf{Tag} & \textbf{Example} \\
    \midrule
    Grammar & Grammatical errors & Which is \{godmother\} of \{Camillo Benso di Cavour\}, whose \{craft\} is \{politician\} ? \\
    & Poor flow/word ordering & Who lives in Anita Bryant whose arrondissement is Pittsburg County? \\
    & Non-idiomatic & What is character role of Turandot ? \\
    \midrule
    Form & Quizlike & astronaut gerhard thiele is associated with which space agency? \\
    & Imperative & find beaufort wind force whose wave height is 0.1 \\
    & Inconcise & Which is the regression analysis that is used by the logistic regression analysis and contains the word logistic in it's name? \\
    
    \midrule
    Meaning & Inconsistent domains/categories & Was 6063 jason invented in eugene merle shoemaker \\
    & Overly specific & Which university attended by arturo macapagal was also the alma mater of hector tarrazona ? \\
    & Redundant constraint &  What is the death place of the \'{e}tienne p\'{e}labon and is the birthplace of the abeille de perrin?\\
    \midrule
    Answerability & Under-constrained & which organism was born on 1926-06? \\
    & Nonsense/Unintelligible & what routed drug that a marketed formulation that has a reference form of neurontin 250 solution? \\
    \midrule
    Factuality & Two questions & Who was married to Faye Dunaway and when did it end? \\
    & Descriptive answer expected & What is a crescent? \\ 
    \bottomrule
\end{tabular}    
\end{table*}

We select three KGQA datasets to sample NL questions from. We choose KGQA datasets that are recent, large, have complex questions and formal queries. We also choose the datasets so that all of the most common KGs are represented in the formal query bindings. Specifically, we consider the datasets DBNQA$^*$~\citep{Linjordet:2020:ICTIR}, LC-QuAD v2.0~\citep{Dubey:2019:ISWC}, and GrailQA~\citep{Gu:2021:WWW}. We then randomly sample 25 NL questions from each of these datasets. Specifically, the 25 NL questions are respectively sampled from the entire DBNQA$^*$ dataset, and from the train splits of LC-QuAD v2.0 and GrailQA. 

Following the approaches of \citet{Arguello:2021:CHIIR} and \citet{Jorgensen:2020:FDG}, we perform an \emph{open coding} pass to collect impressions on how the NL questions fall short of being ``natural.'' Three academic researchers are presented each NL question and asked to  (i) judge whether or not the question is natural, (ii) produce a (more) natural paraphrase of the question, and (iii) comment on the NL question and suggest any tags or categories regarding ``why and how the question is or is not natural.'' 
The first author then collates the responses, and the comments and categories are harmonized into a consistent coding scheme of tags by the first author. Both authors review the extracted tags and discuss common themes across tags.  The tags are then organized into the five dimensions of unnaturalness illustrated along with NL question examples in Table~\ref{tab:coding}. We note that the examples in the Table may exhibit more than one of the properties the exemplify a given tag or dimension of unnaturalness. 
\begin{figure*}[!h]
   \centering
   \begin{tabular}{c@{}c@{}c@{}}
   \multicolumn{3}{c}{\textbf{All NL questions}}\\
     \includegraphics[width=.32\textwidth]{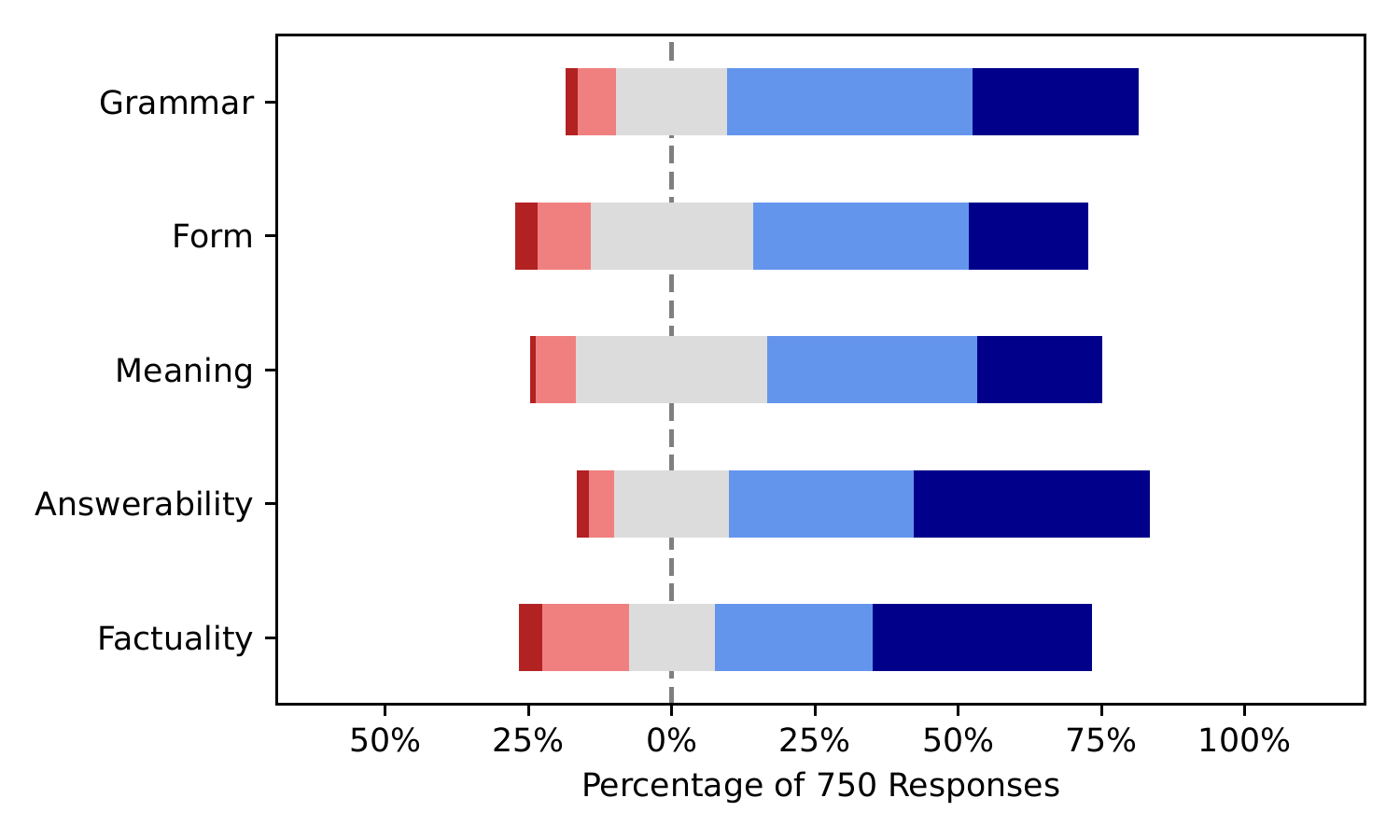} & 
     \includegraphics[width=.32\textwidth]{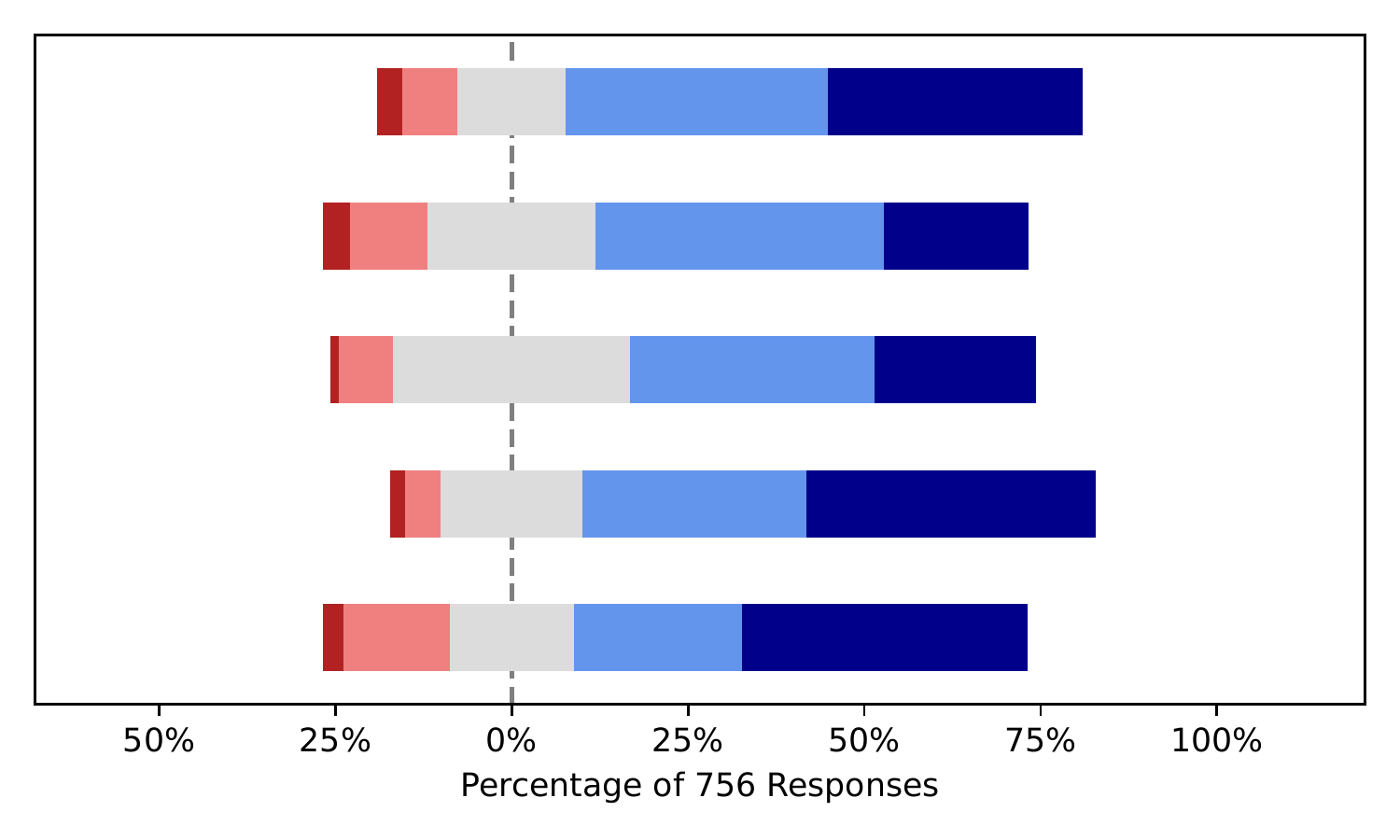} & 
     \includegraphics[width=.32\textwidth]{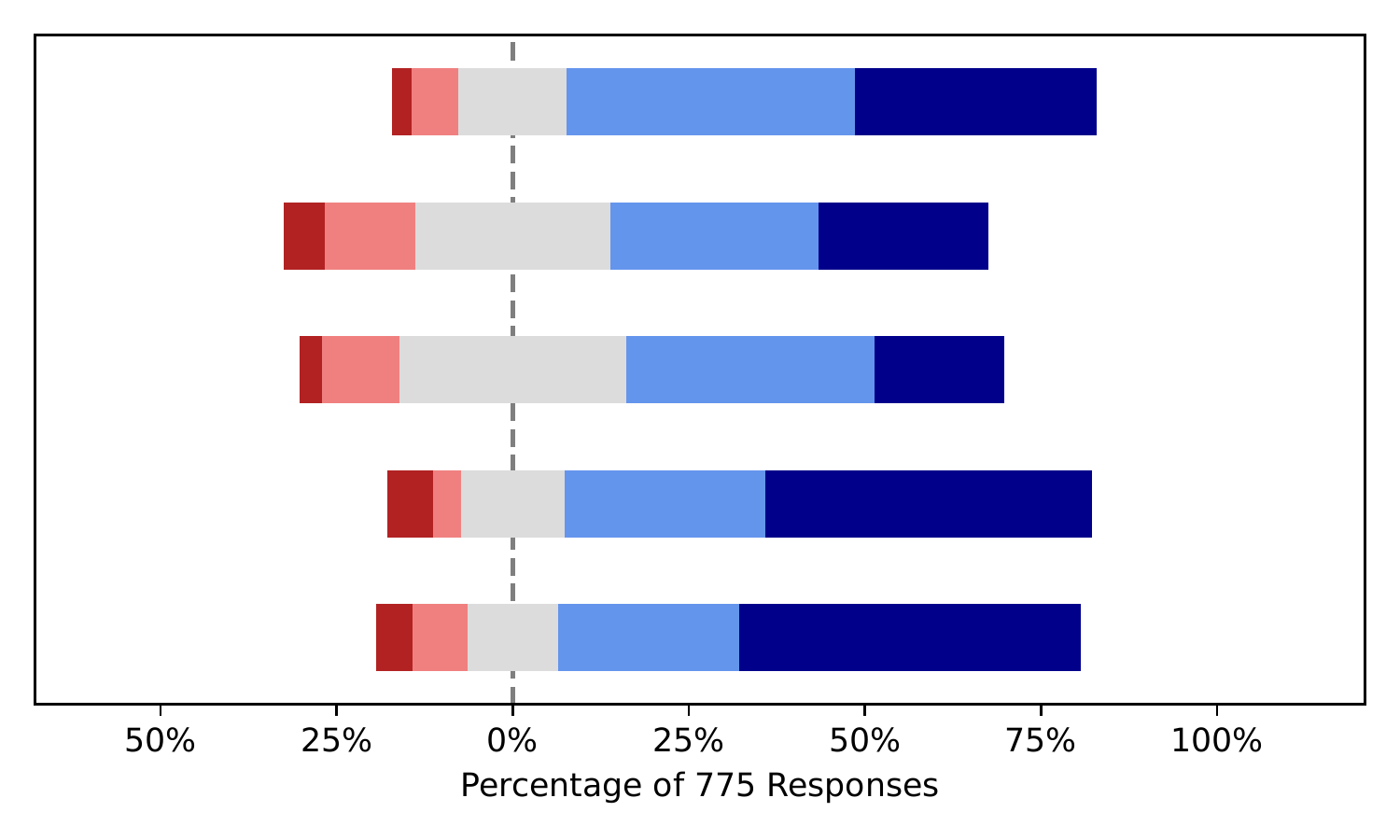}
     \\  
   \multicolumn{3}{c}{\textbf{Questions rewritten}}\\
     \includegraphics[width=.32\textwidth]{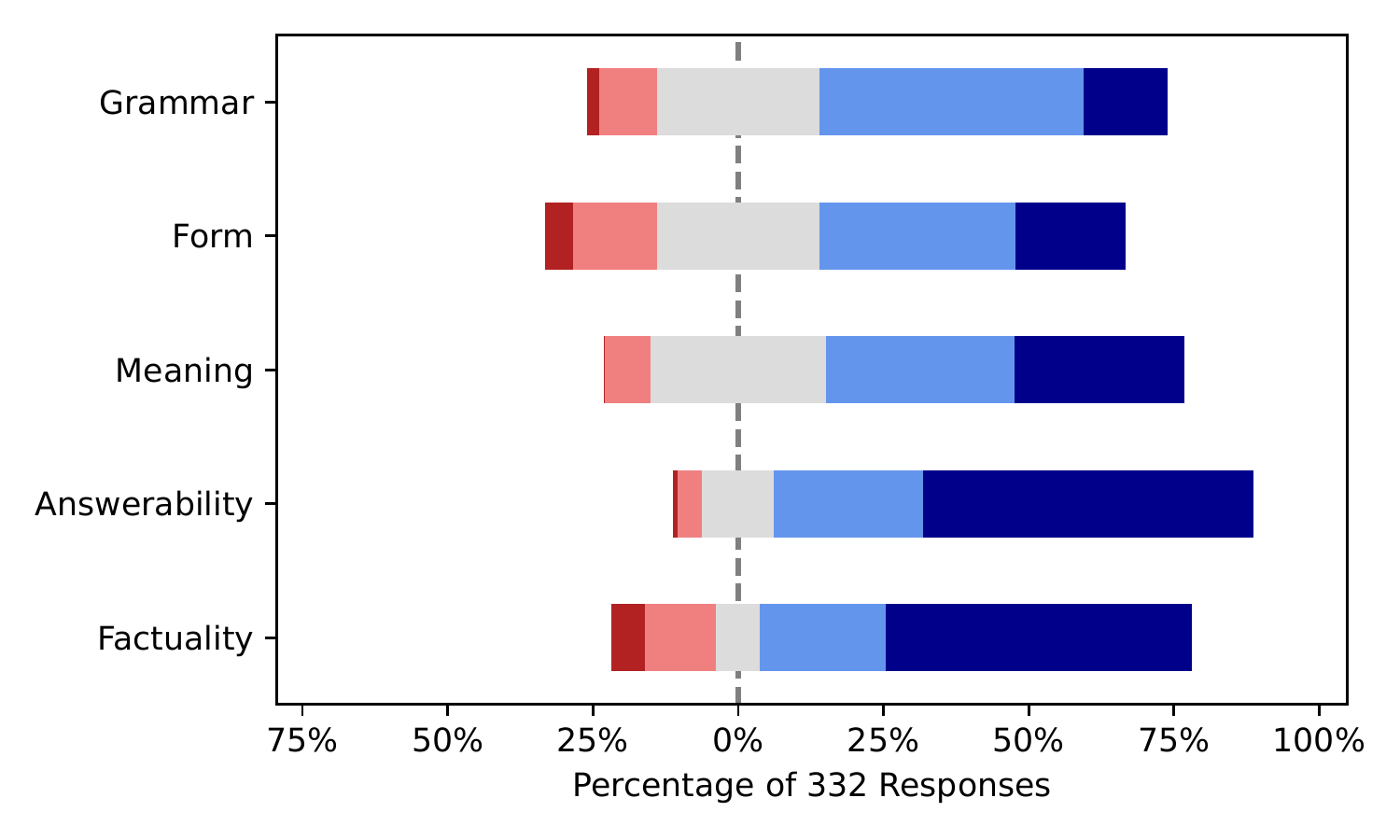} & 
     \includegraphics[width=.32\textwidth]{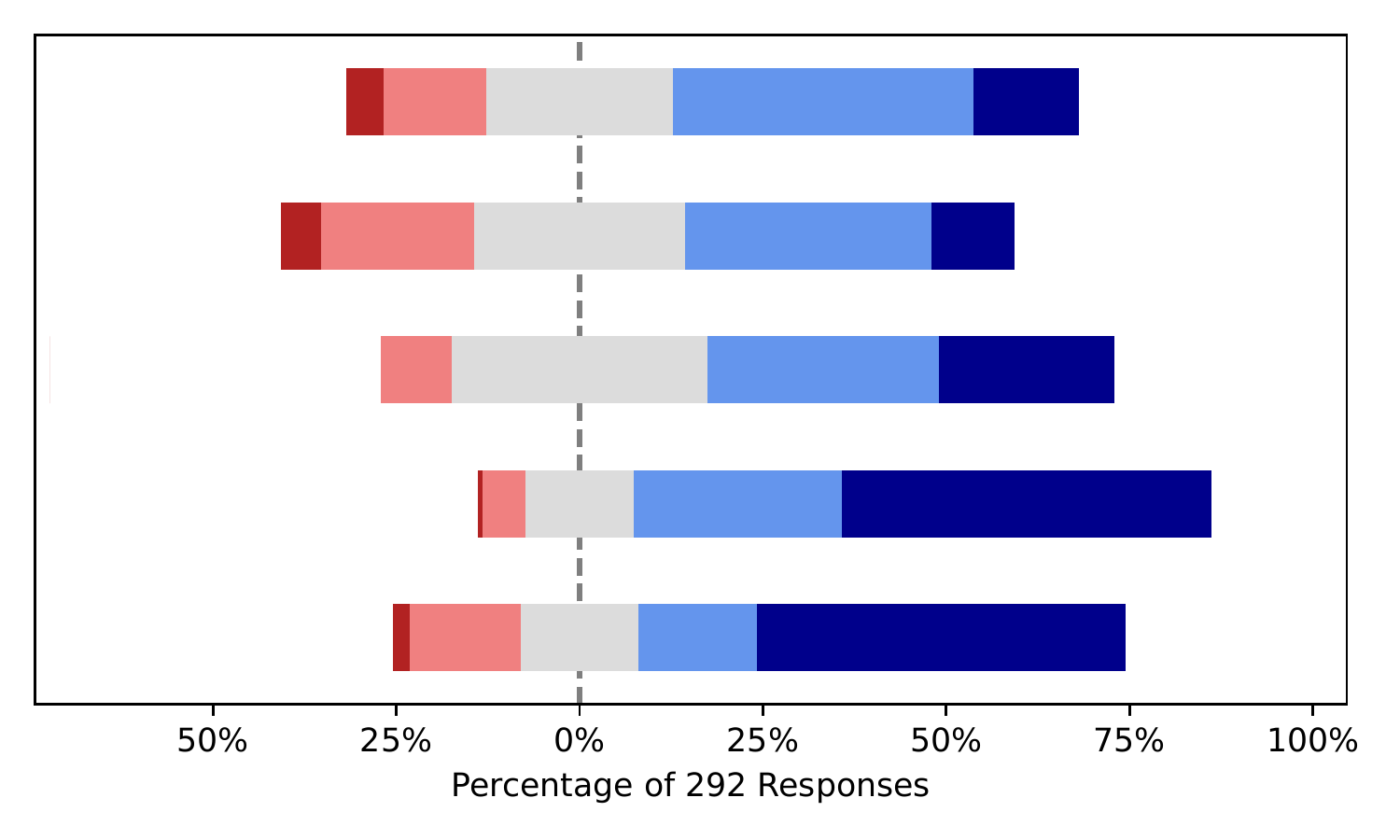} & 
     \includegraphics[width=.32\textwidth]{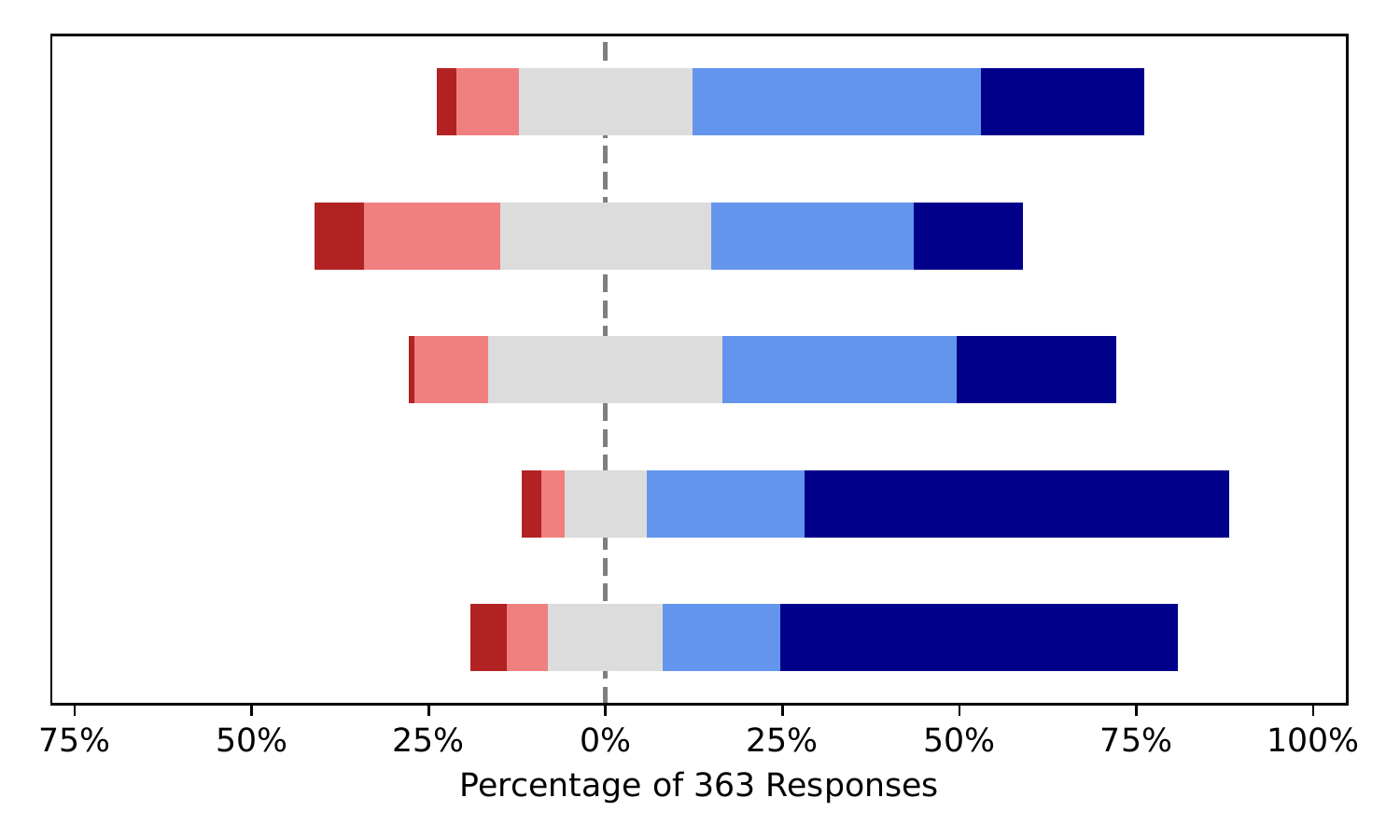}
     \\
   \multicolumn{3}{c}{\textbf{Questions not rewritten (``already perfect'')}}\\
     \includegraphics[width=.32\textwidth]{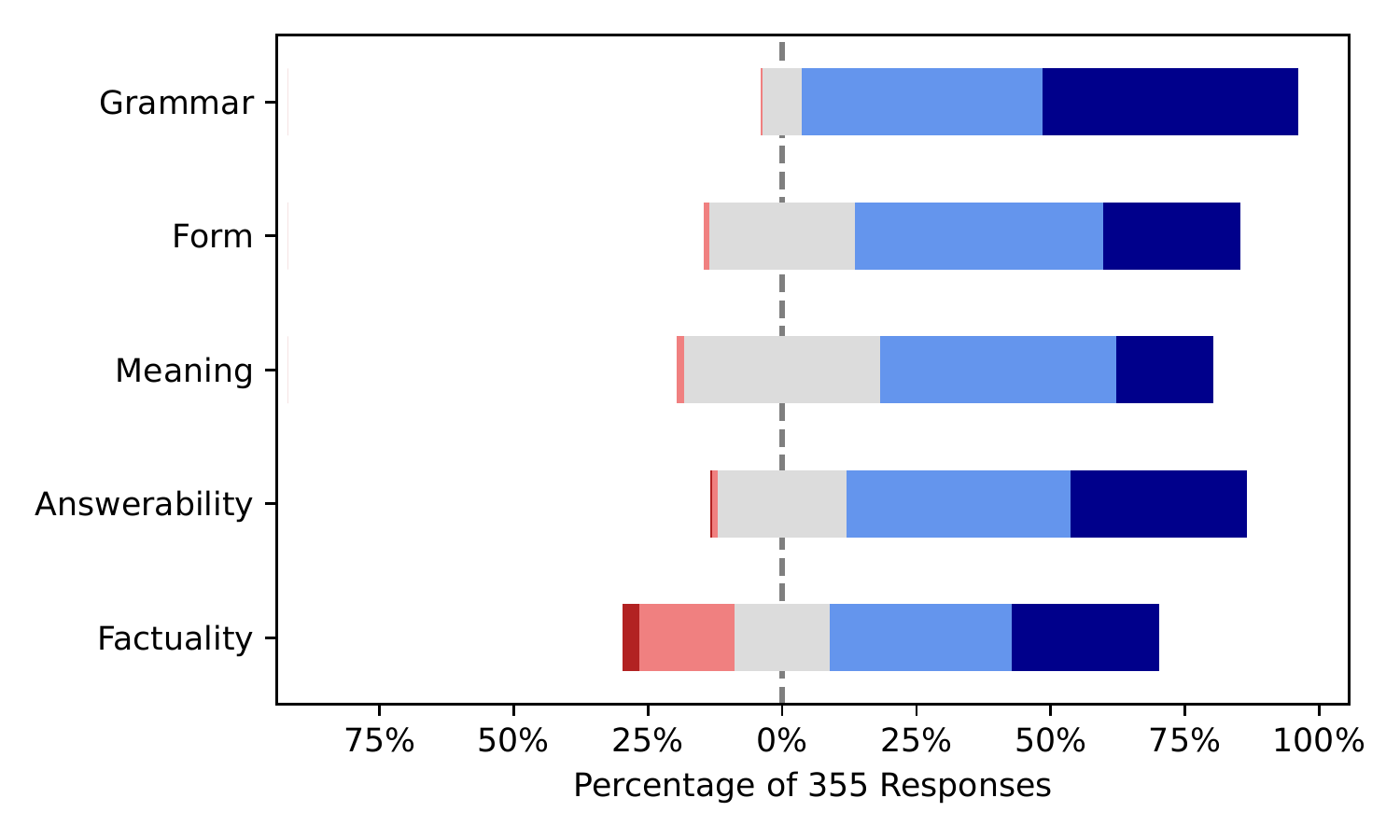} & 
     \includegraphics[width=.32\textwidth]{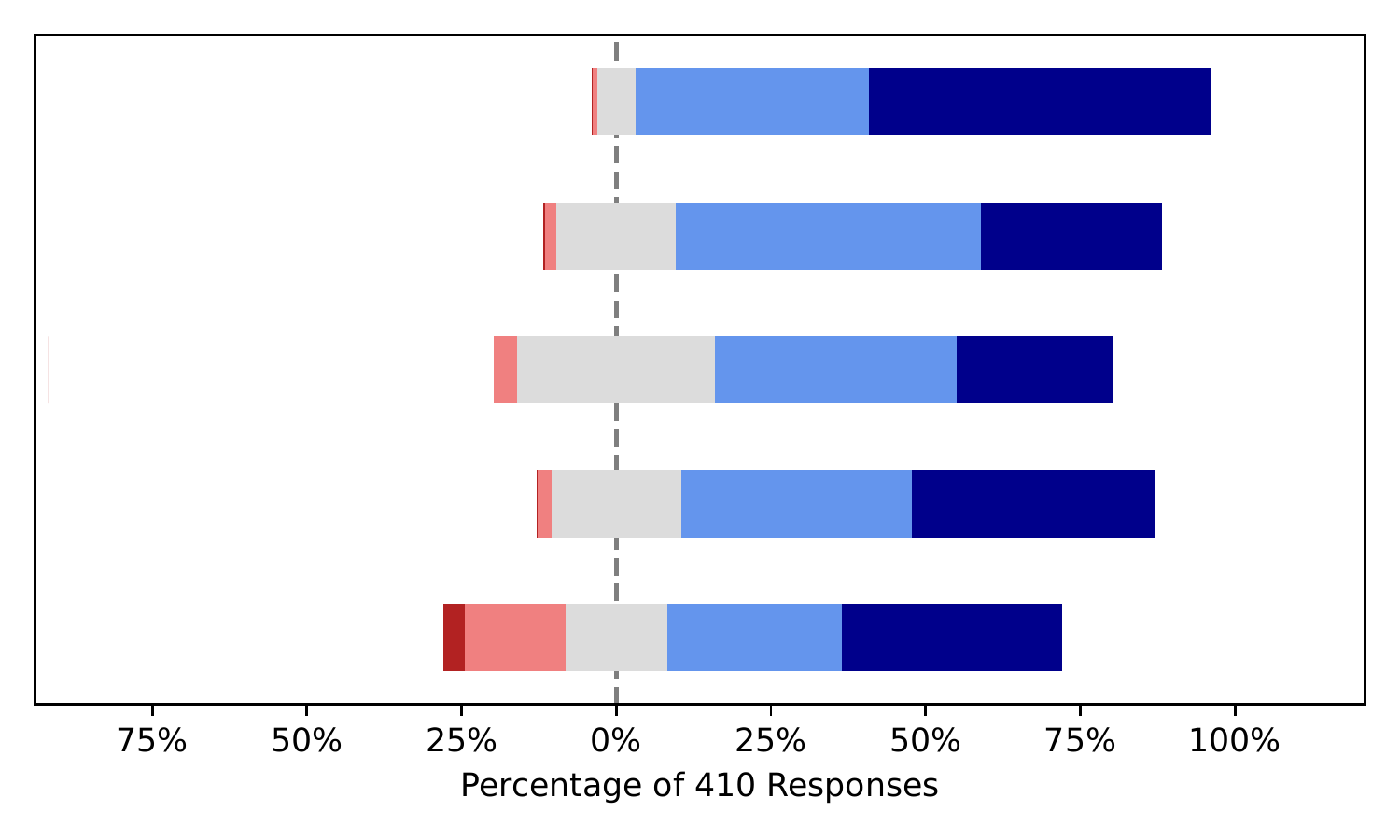} & 
     \includegraphics[width=.32\textwidth]{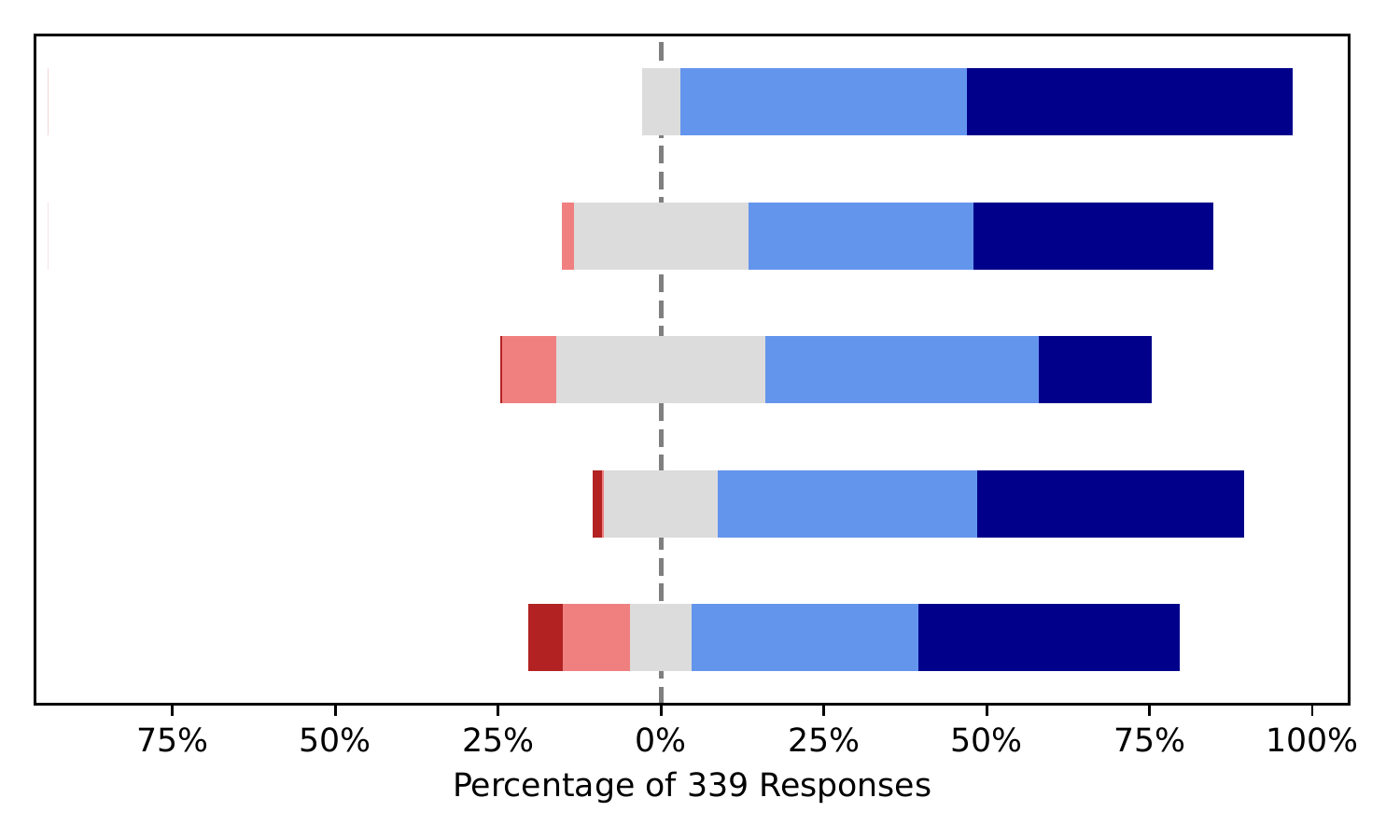}
     \\    
   \multicolumn{3}{c}{\textbf{Questions not rewritten (``unclear'')}}\\
     \includegraphics[width=.32\textwidth]{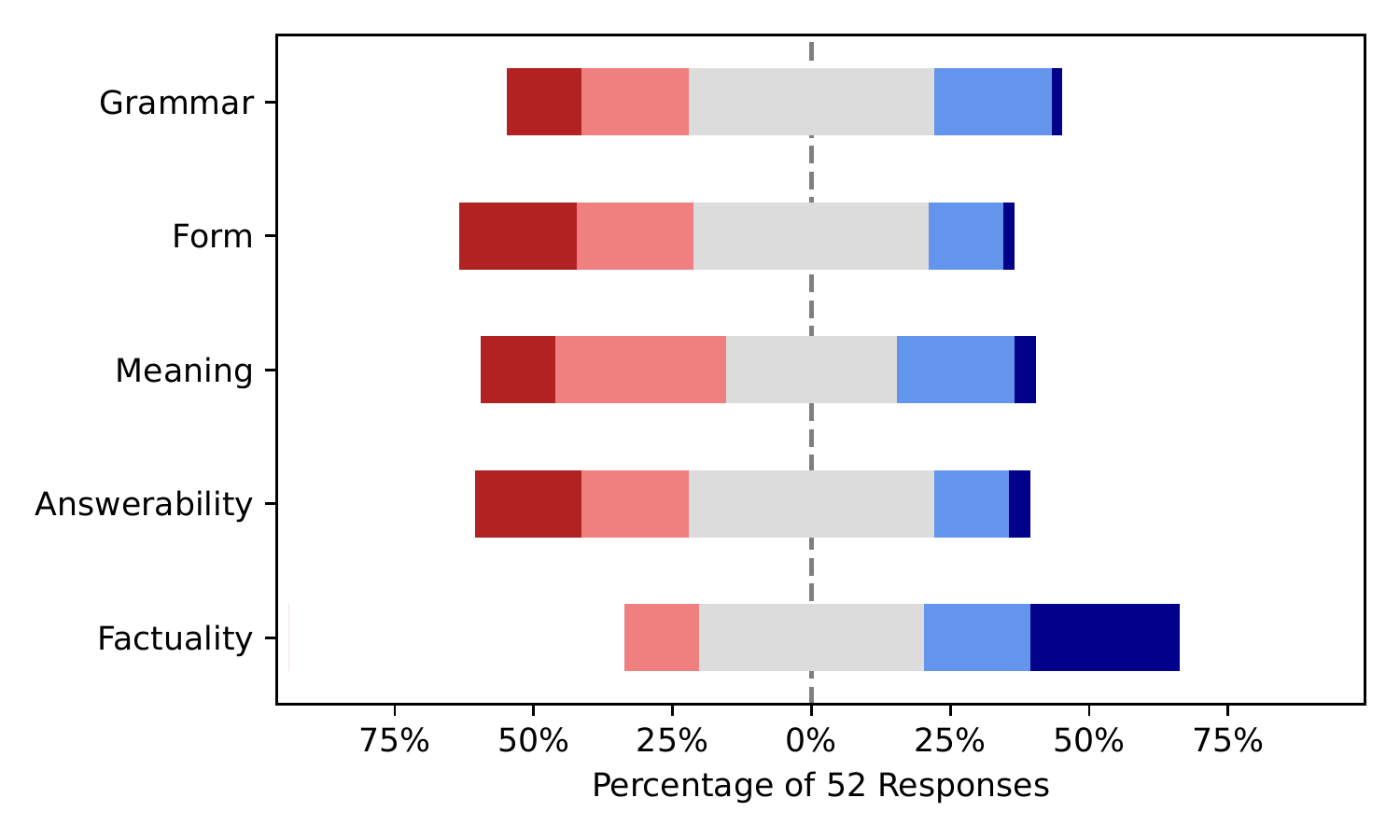} & 
     \includegraphics[width=.32\textwidth]{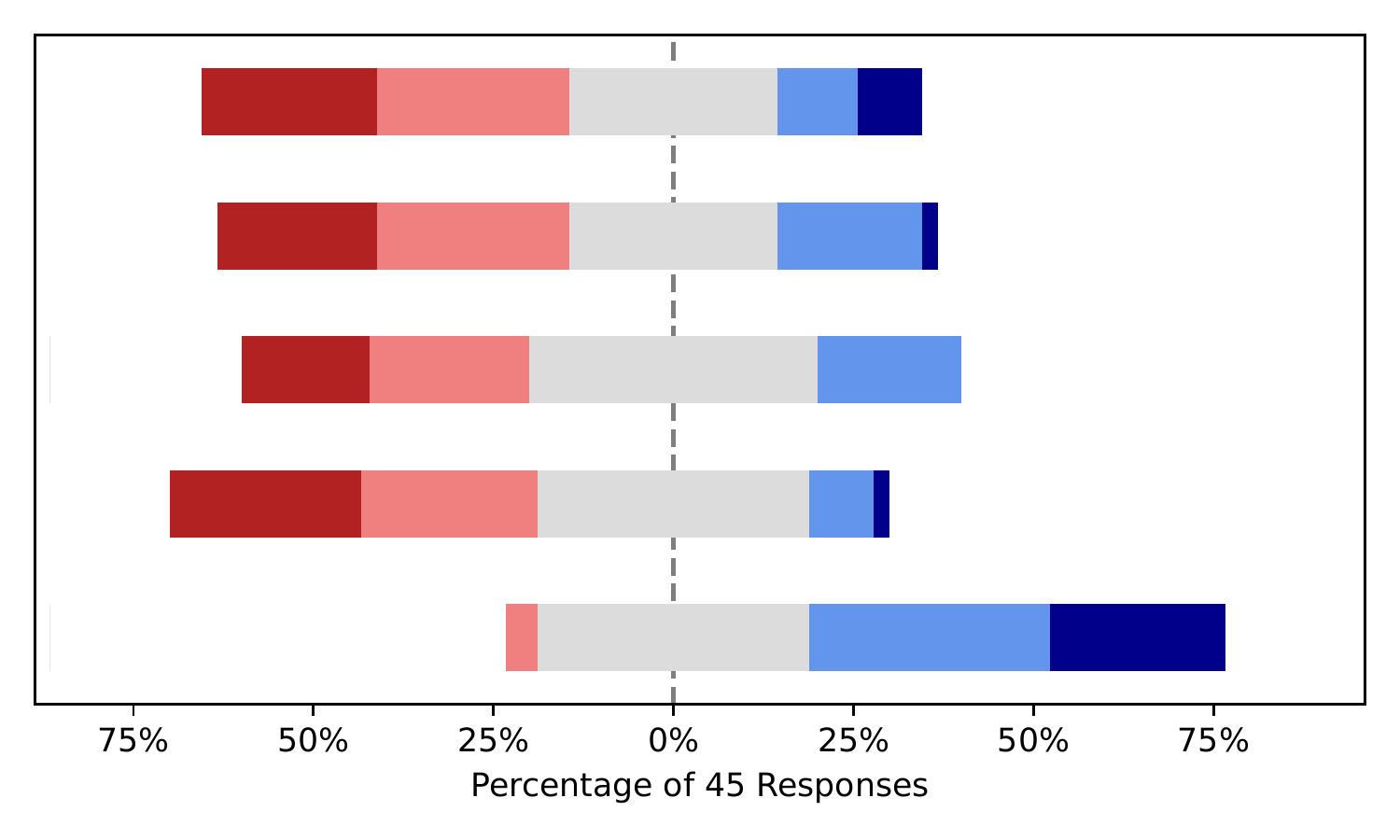} & 
     \includegraphics[width=.32\textwidth]{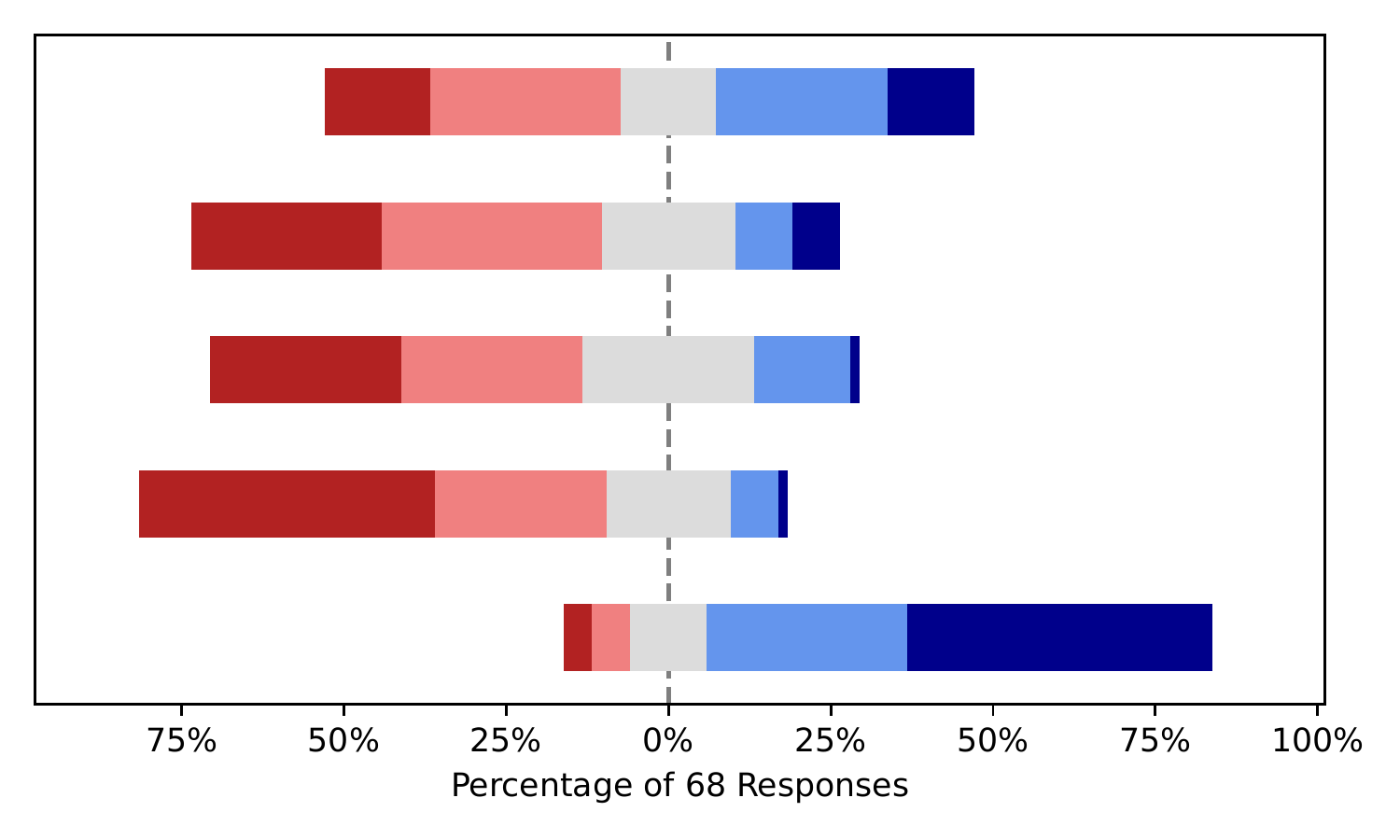}
     \\    
   \multicolumn{3}{c}{\textbf{Questions not rewritten (another reason)}}\\
     \includegraphics[width=.32\textwidth]{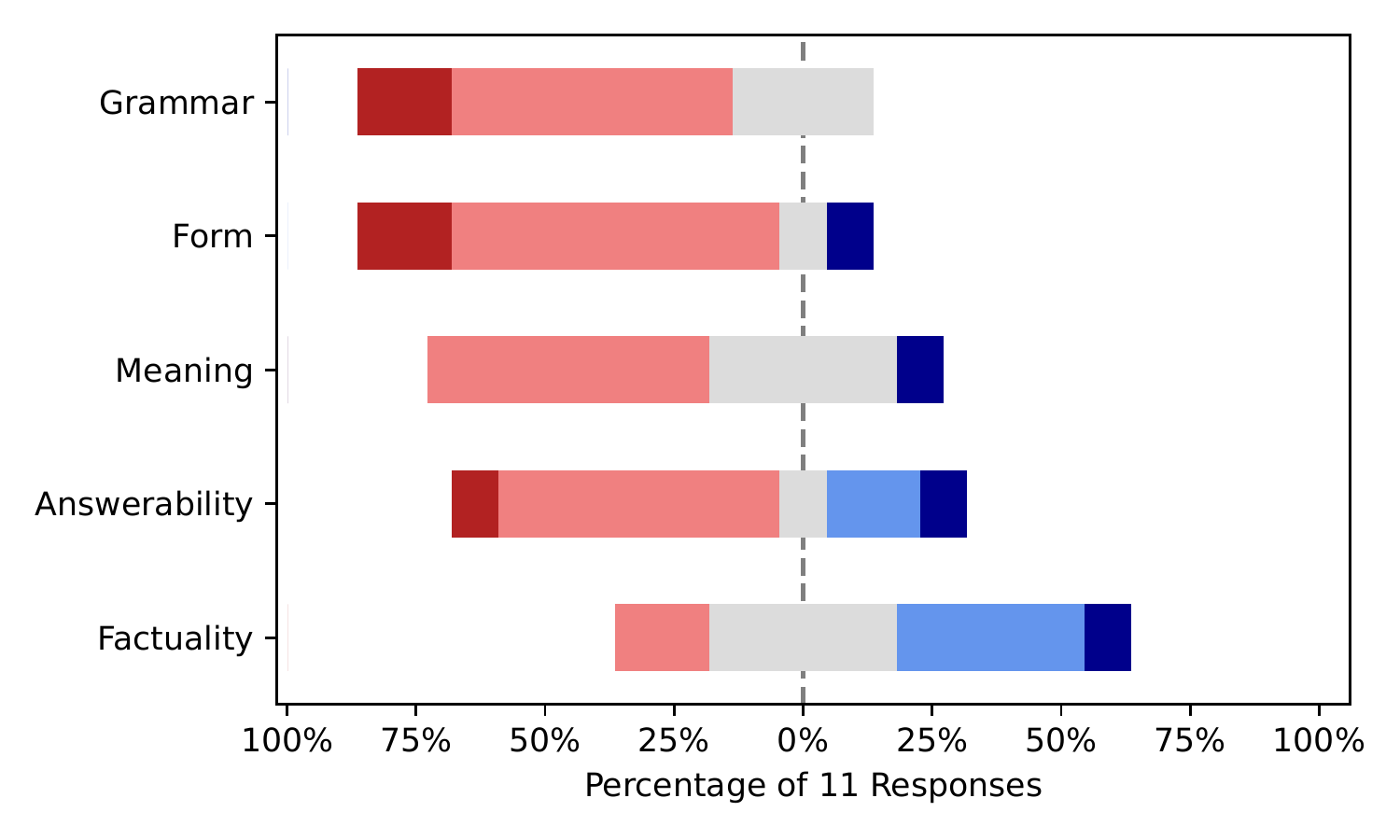} & 
     \includegraphics[width=.32\textwidth]{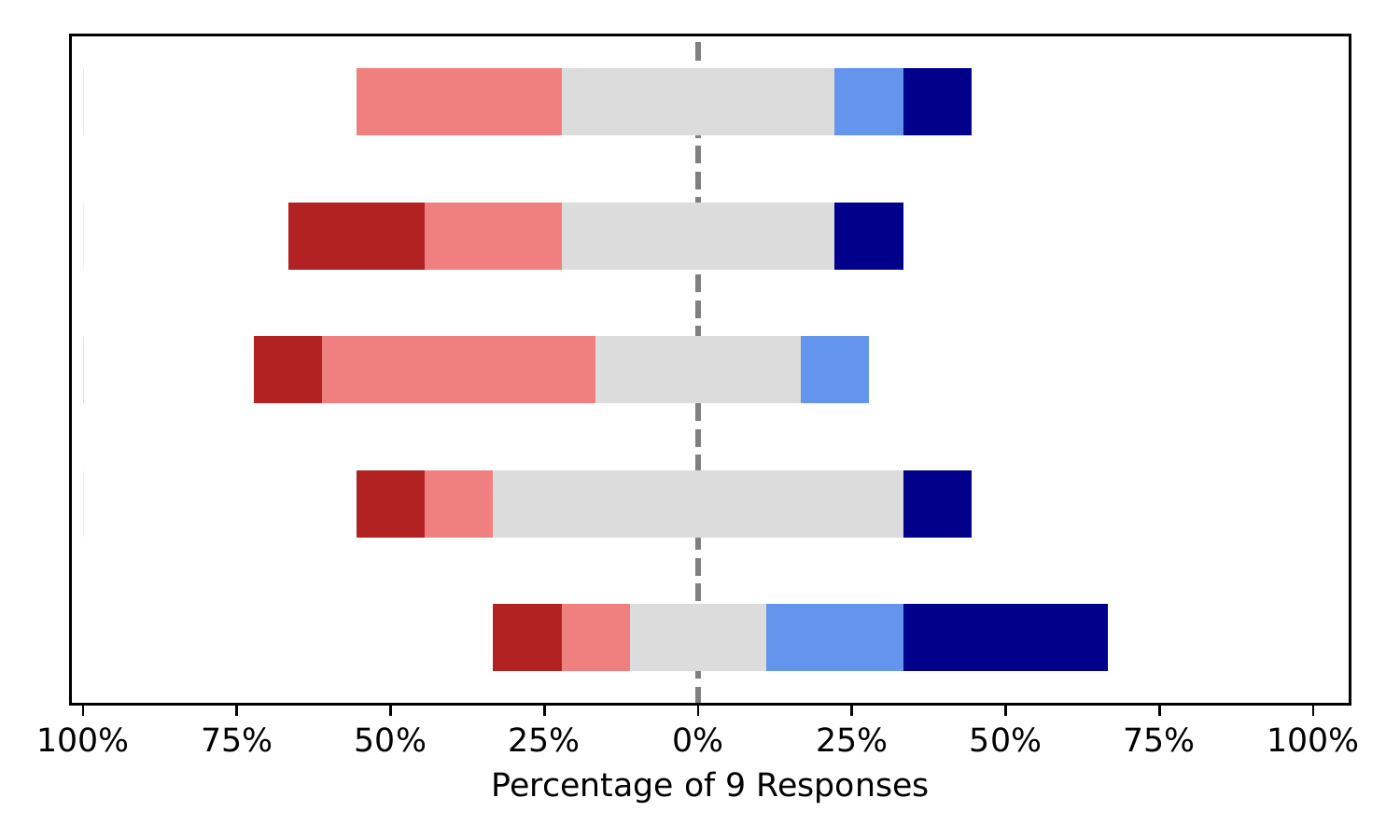} & 
     \includegraphics[width=.32\textwidth]{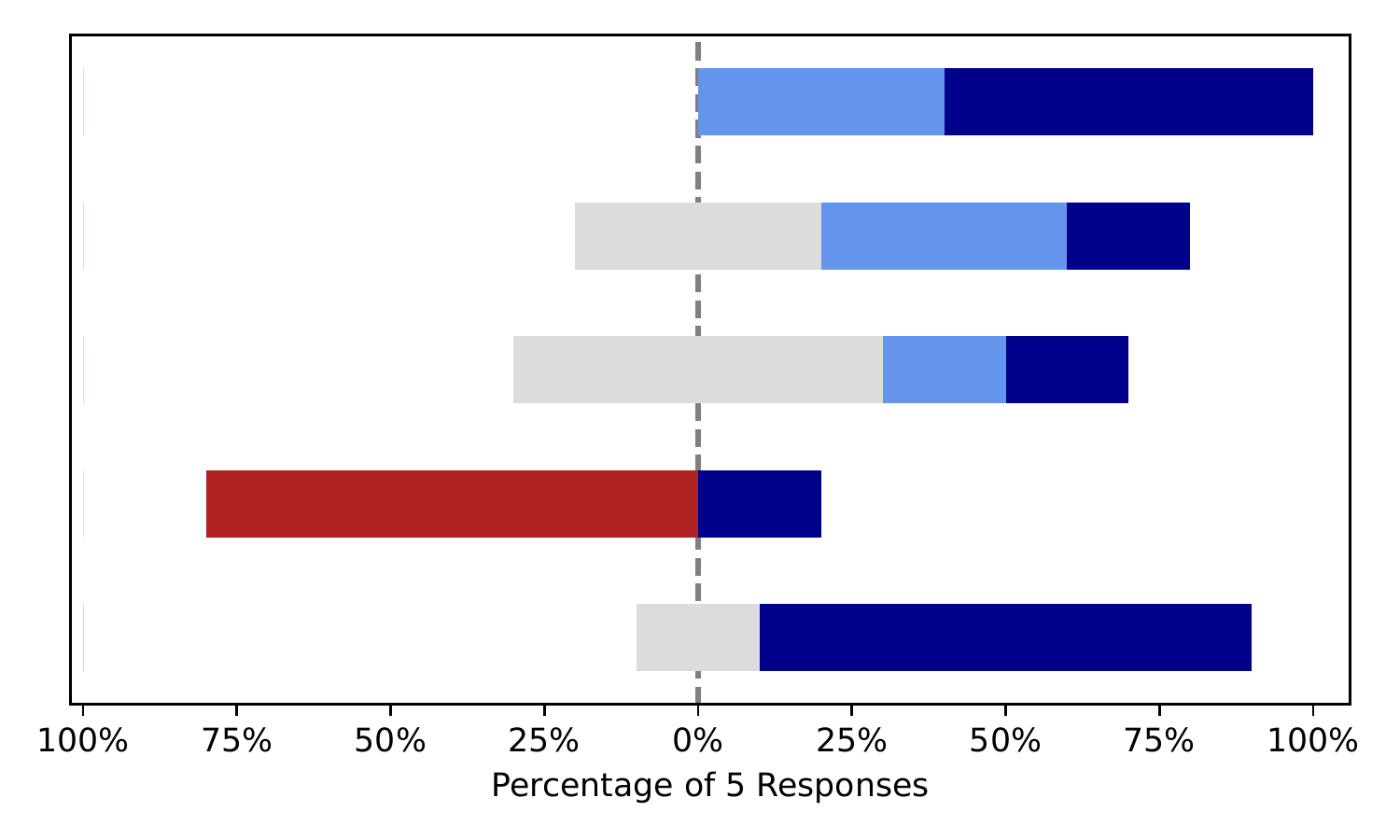}
     \\ 
   DBNQA$^*$ & LC-QuAD v2.0 & GrailQA \\  
   \end{tabular}    
\caption{Likert scale rating results on all NL questions (Top), on questions which were rewritten (Second row), on questions which were not rewritten because the original question was ``already perfect'' (Middle), on questions which were ``unclear'' (Fourth row), and on questions which were not rewritten for another reason (Bottom), broken down per dataset (columns).}
   \label{fig:likert}
\end{figure*}

\section{Data Annotation}
\label{sec:annotation}

Having defined codes to characterize question unnaturalness in KGQA datasets, we next design a protocol for larger-scale data labeling using a two-step crowdsourcing pipeline.  
Crowd workers are first asked to annotate and paraphrase the sampled NL questions.  Then, in a separate task, a different set of workers is employed to select the best version of a question from a set, including the original formulations as well as rewritten questions from the first task. 

We sample a new set of NL questions, this time 250 NL questions from each of the three datasets. Specifically, we randomly sample the NL questions from the test split of DBNQA$^*$ and LC-QuAD v2.0, but from the validation split of GrailQA, since for the latter, the public test split does not include ground truth answers. 
The resulting test collection is termed IQN-KGQA and is summarized in Table~\ref{tab:iqn_kgqa}.

\subsection{Crowdsourcing: Platform and Workers}

Our data annotation was conducted on the Amazon Mechanical Turk platform. 
For both tasks, workers were required to have a HIT approval rate of 98\% with more than 1000 approvals. The payments were set to USD \$0.30 and \$0.15, respectively, based on the estimated effort demanded for each task.

Crowd workers were not required to have domain knowledge, based on the findings of \citet{Dubey:2019:ISWC}. Since only open-domain KGQA datasets are used in the present work, the annotation tasks are designed to rely on common sense and English language knowledge primarily. For example, the prompt for the Likert scale ``answerablility'' is the question ``Would you be able to answer this question with the help of a search engine or Wikipedia?'' In other words, the data annotation relies on metacognition with respect to an NL question rather than actually finding some answer. Identifying crowd workers with comparable levels of expertise in specific domains prior to data annotation would present a major additional cost.  Also, those workers would not necessarily be representative of the general user population whose information needs KGQA datasets aim to capture.

\begin{table}[t]
\caption{Summary of the IQN-KGQA collection.}
\label{tab:iqn_kgqa}
\centering
\begin{tabular}{llrr} 
    \toprule
    \textbf{Subset} & \textbf{Split} & \textbf{\#Questions} & \textbf{\#Rewritten} \\
    \midrule
    DBNQA$^*$ & test & 250 & 180 \\
    LC-QuAD v2.0 & test & 250 & 150 \\
    GrailQA & validation & 250 & 211 \\
    \midrule
    Total & & 750 & 541 \\
    \bottomrule
\end{tabular}    
\end{table}

\subsection{Task 1: Annotate and Rewrite}

In the first task, the crowd workers are given one of the sampled NL questions (the target question) and are asked to rate the question in terms of the five dimensions of unnaturalness. For each dimension, the question is rated on a Likert scale. Next, the crowd workers are asked to rewrite the question, to ``write a better, more natural and correct version'' of the question. Finally, the crowd workers are asked to indicate if they rewrote the question, and if not what the reason was, including a free text field to elaborate on any ``other'' reason for not rewriting. The complete form and instructions are provided in the GitHub repository accompanying this paper.

The responses from crowd workers are then quality controlled, and responses which overtly demonstrate a lack of genuine effort are entirely removed. Criteria for this exclusion include indicating that a question was rewritten but providing no paraphrase, writing a short comment like ``good'' instead of a question, or else copy-pasting parts of the instructions into the rewrite field.  

Whenever crowdsourced responses are excluded, additional responses are requested, so that every sampled NL question is annotated (and potentially rewritten) with acceptable responses by at least three different crowd workers. 

The results of the Likert scale ratings are shown in Fig.~\ref{fig:likert}. The scales are oriented so that the farther to the right the scale lies, the more natural the questions are considered by the crowd workers. The differences between the rows are intuitive since the rows group the responses in terms of the reason given for whether the original question has been rewritten or not. Specifically, the middle row reflects responses where the crowd worker deems the original question to be ``already perfect'' and hence abstains from rewriting the question. This is also the row with the highest ratings over all five Likert scales. 

The bottom two rows also show a consistency between the Likert scale ratings and the reason given why the original question was not rewritten. However, here the ratings are mostly negative compared to the distribution over all responses. The bottom two rows' ratings are also negative compared to the second and middle rows from the top, which reflect the original question having been rewritten or being ``already perfect.'' One interesting exception is that the Likert scale ``factuality'' is rated highly even in the bottom two rows of Fig.~\ref{fig:likert}. It is possible that the distinction was not made clear to the crowd workers between whether a question indicates a very terse and factual answer or a longer, more descriptive one. Alternatively, it may be possible for a question to clearly indicate that its proper answer is factual, but that what is asked is so unclear that the question cannot be improved. 

Overall, the consistency of ratings across the five Likert scales over the 750 sampled NL questions as rated in the approved crowdsourced responses is calculated as Cronbach's $\alpha$ = 0.707, which is designated as ``acceptable.''

\subsection{Task 2: Validate and Vote}

We then use the rewritten questions from the previous task to establish which version of an NL question is the better formulation. For every original question where at least one rewrite was provided by crowd workers in Task 1, we take the original question and up to three rewrites, shuffle the order, and ask a different set of crowd workers to choose which version of the question is the best way of asking.  See the GitHub repository for the specific instructions given to crowd workers.

For this task, since the response type is very simple and if less than three rewritten questions were generated there is always at least one non-option which the crowd worker technically can choose, quality control consists of removing responses from crowd workers who repeatedly choose non-options. 

\begin{figure}[t]
   \centering
   \includegraphics[width=0.4\textwidth]{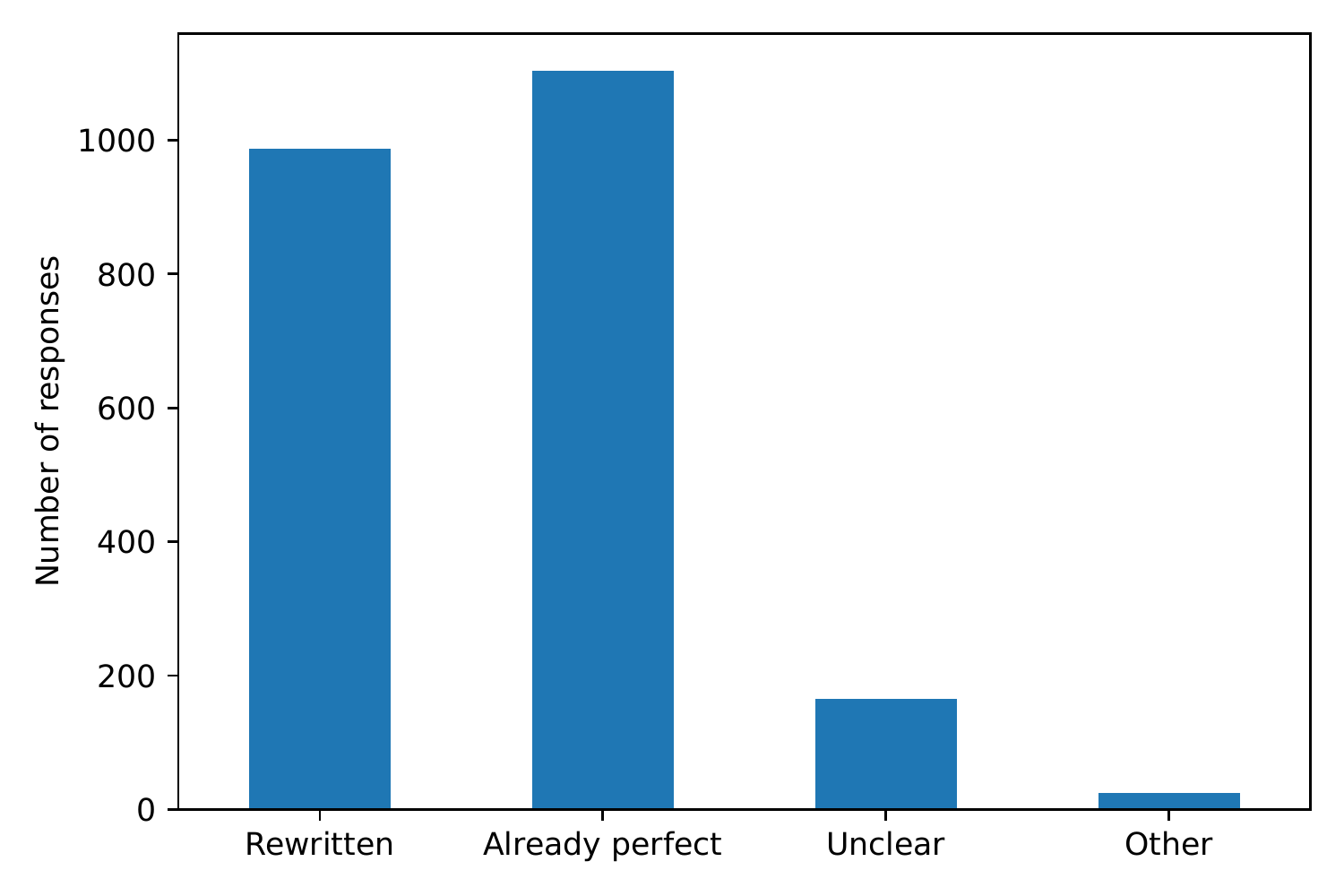}
\caption{Histogram of frequencies over whether a rewritten question was provided (``rewritten'') or otherwise reasons given for not rewriting (the original question was ``already perfect'' or ``unclear,'' or else some ``other'' reason).}
   \label{fig:reasons_histogram}
\end{figure}

\begin{figure}[t]
   \centering
   \includegraphics[width=0.4\textwidth]{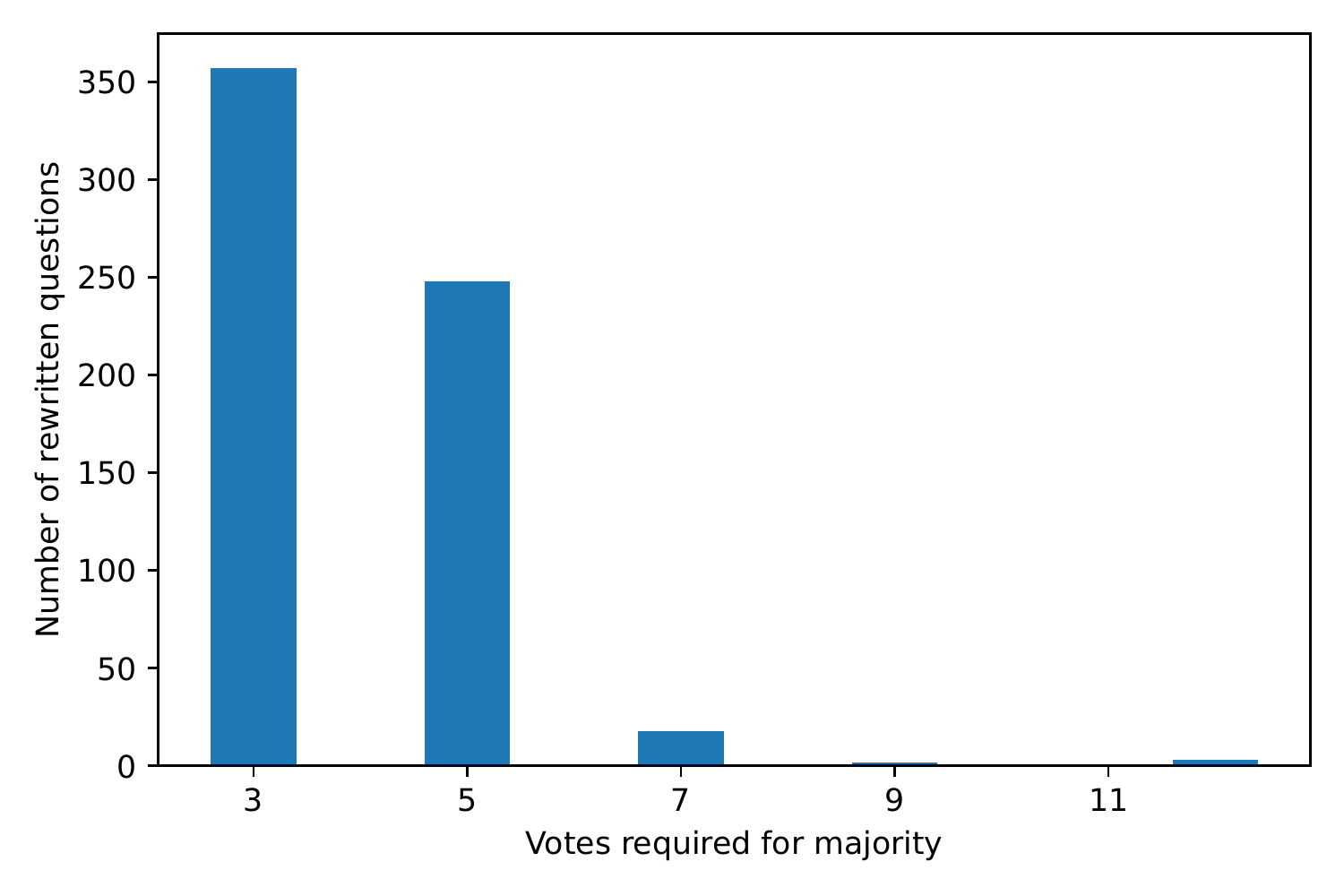}
\caption{Histogram of frequencies over how many votes were required to determine a majority in favour of one version of the original or rewritten question. 
}
   \label{fig:majority_votes}
\end{figure}

Each question and its rewrites are validated by at least three crowd workers. If the Task 2 result is a clear majority for any specific version of the question, then that is the question carried forward into the rewritten questions test collection. If there is not a clear majority given a set of original question and its rewrites, two more crowd worker validations are requested until a majority vote emerges. The distribution of responses is displayed in Fig.~\ref{fig:reasons_histogram}, while the number of crowd worker responses required to reach a majority is shown in Fig.~\ref{fig:majority_votes}.  
In total, 541 of the 750 questions are rewritten in the new collection; see Table~\ref{tab:iqn_kgqa} for a breakdown on specific subsets.
\section{Experiments}
\label{sec:experiments}

We compare model performance on the original versus rewritten NL questions in our samples.  Specifically, we use neural KGQA models trained on the DBNQA$^*$~\citep{Linjordet:2020:ICTIR} and GrailQA~\citep{Gu:2021:WWW} datasets.\footnote{Since we could not find papers with open source code addressing LC-QuAD v2.0 and there are still no models on the corresponding leaderboard, this dataset is not included in our experiments.} 
%Our the results demonstrate that with additional quality improvements on the NL questions the trained models actually perform worse. => let's phrase it as a question instead, as at this point we don't know
The question we seek to answer is how quality improvements on the input NL questions impact the answer prediction effectiveness of the models.  

\subsection{Experimental Setup}

For each of the KGQA datasets, the underlying KG is provided by a Virtuoso triplestore instance. For DBNQA$^*$, DBpedia 2016 is the KG used to execute formal queries to retrieve answers. For GrailQA, Freebase is served as the KG, following the instructions provided by~\citet{Gu:2021:WWW}.\footnote{\url{https://github.com/dki-lab/Freebase-Setup}} This includes using their processed version of Freebase to make it fully compatible with the relevant Resource Description Framework (RDF) standard.  

The GrailQA models rely on entity linking, which is provided for the full GrailQA validation split. In order to compare the original and rewritten question samples under equivalent conditions, the rewritten questions are identified with the original questions' query ID to apply the same entity linking to the rewritten NL question. 

\begin{table*}[t!]
\caption{Results on the DBNQA* and GrailQA datasets. The full subset refers to the original benchmarks and is included for reference.  The IQN-KGQA dataset contains a sample of 250 questions per dataset.  Performance is reported on the original questions in the sample as well as on their rewritten variant with improved naturalness. Significance is tested between the rewritten and original questions.}
\label{tab:results}
\vspace*{-0.5\baselineskip}
\centering
\begin{tabular}{lllccccccc} 
    \toprule
    \textbf{Dataset} & \textbf{Subset} & \textbf{Method} & \multicolumn{2}{c}{\textbf{Full subset}} & & \multicolumn{4}{c}{\textbf{IQN-KGQA}} \\ 
    \cline{4-5} \cline{7-10}
    & &&  \multicolumn{2}{c}{\textbf{Original questions}} & & \multicolumn{2}{c}{\textbf{Original questions}} & \multicolumn{2}{c}{\textbf{Rewritten questions}} \\
    & & & \textbf{EM} & \textbf{F$_1$} & & \textbf{EM} & \textbf{F$_1$} & \textbf{EM} & \textbf{F$_1$} \\
    \midrule
    DBNQA$^*$ & Test & NSpM baseline & 0.000 & 0.013 & & 0.000 & 0.020 & 0.004 & 0.016 \\
    & & NSpM+Att1 & 0.081 & 0.119 & & 0.085 & 0.105 & 0.033$^\dag$ & 0.050$^\dag$ \\
    & & NSpM+Att2 & 0.089 & 0.132 & & 0.081 & 0.117 & 0.028$^\dag$ & 0.050$^\ddag$ \\
     & & ConvS2S & 0.091 & 0.138 & & 0.121 & 0.152 & \textbf{0.036}$^\ddag$ & 0.048$^\ddag$ \\
     & & Transformer & \textbf{0.177} & \textbf{0.260} & & \textbf{0.166} & \textbf{0.254} & \textbf{0.036}$^\ddag$ & \textbf{0.067}$^\ddag$ \\
    \midrule
    GrailQA & Dev & Ranking+BERT & \textbf{0.510} & \textbf{0.583} & & \textbf{0.452} & \textbf{0.540} & \textbf{0.372} & \textbf{0.452}$^\dag$ \\
    & & Transduction+BERT & 0.337 & 0.364 & & 0.296 & 0.339 & 0.208$^\dag$ & 0.251$^\dag$ \\
    \bottomrule
\end{tabular}    
\end{table*}

\subsection{Methods}

We use seven different neural KGQA methods to test the effect of rewritten NL questions on KGQA performance.  These are sequence-to-sequence neural models with an encoder-decoder motif, where all but one are used to generate the formal query as a sequence of tokens. The exception is Ranking+BERT~\citep{Gu:2021:WWW}, where a neural model is used as a ranker to rank generated candidate formal queries.

% \todo{Would be nice to list the models as item lists with the model name boldfaced. There can be some narrative text in between the items, discussing  settings, etc. that apply to the group.}

Three methods are variations of the Neural Sparql Machine (NSpM)~\citep{Soru:2017:arXiv, Soru:2018:ICML-ws, Yin:2021:FGCS} architecture, including the NSpM baseline, NSpM+Att1, and NSpM+Att2 models. They are all based on Tensorflow NMT. NSpM+Att1 features a normed Bahdanau~\citep{Bahdanau:2014:arXiv} attention mechanism, while NSpM+Att2 uses a scaled Luong~\citep{Luong:2015:arXiv} attention mechanism. All three NSpM models are specified with 2 layers and a dropout coefficient of 0.2. They are also all trained for 50,000 training steps. 

Two methods, ConvS2S~\citep{Gehring:2017:ICML} and Transformer~\citep{Vaswani:2017:NeurIPS}, are adapted from machine translation between natural languages to semantic parsing for KGQA.
We rely on the sequence-to-sequence model implementations in Pytorch.\footnote{\url{https://github.com/bentrevett/pytorch-seq2seq/}} 
To better support the models, the both NL and formal query data is pre-processed with sub-word tokenization, specifically Byte Pair Encoding (BPE)~\citep{Sennrich:2015:arXiv} using Sentencepiece.\footnote{\url{https://github.com/google/sentencepiece}} The other hyperparameters for ConvS2S and Transformer are kept as default, except notably the training data is not shuffled between epochs during training, and the models are trained in a case-sensitive manner.

The five models mentioned thus far are all trained on the training split of DBNQA$^*$, which has been shuffled. The models predict formal queries from the NL questions in the full test split of DBNQA$^*$, as well as the original sample of 250 NL questions from the test split, and the rewritten NL questions of the same sample. 

Next, we use the methods using a pre-trained model based on BERT~\citep{Devlin:2019:NAACL-HLT} provided by~\citet{Gu:2021:WWW}.\footnote{\url{https://github.com/dki-lab/GrailQA/}} Specifically, the model is an LSTM-based sequence-to-sequence which uses uncased base-BERT for encoding, and is fine-tuned on GrailQA train split. The Transduction+BERT method uses this model for generating a formal query in an auto-regressive manner. In contrast, Ranking+BERT uses this model to rank candidate formal queries. 

% In addition setting up the KG backends as appropriate for the datasets, the KGQA models differ in their need for pre-processing of the instances. 

\subsection{Results and Analysis}

With the methods described above, we achieve the KGQA performance results listed in Table~\ref{tab:results}. We use two performance measures, Exact Match (EM) and F$_1$, to quantify effectiveness. Exact match compares the predicted formal query to the ground truth formal query. For DBNQA$^*$, EM is 1.0 for an instance if and only if the two strings are identical. Meanwhile, using the provided evaluation script with GrailQA~\citep{Gu:2021:WWW}, the predicted and ground truth formal queries are both converted to query graphs and are considered as exactly matching if the graphs are isomorphic. The F$_1$ measure is based on the precision and recall of comparing answer sets. For the KGQA models evaluated on DBNQA$^*$, if both the ground truth answer and the predicted answer are empty sets, the score for an instance is 1.0. This follows the example of \citet{Usbeck:2019:SW}.   

In Table~\ref{tab:results}, we observe that the original questions in our sample (IQN-KGQA) may differ in terms of mean performance when compared to the full subset from which the sample was taken. The difference can be either either lower (e.g., Transformer) or higher (e.g., ConvS2S) on the sample than on the full subset.
However, performance on original questions are of the same magnitude for both the full subset and sample across all methods. This holds true for both DBNQA$^*$-test and GrailQA-dev.

In contrast, performance is reduced drastically when predicting on the rewritten questions. The three methods with the highest performance overall, Transformer, Ranking+BERT, and Transduction+BERT, all show that performance in both EM and F$_1$ is reduced by a large fraction (up to 78\%) when predicting on the rewritten questions compared to predicting on the original questions. 
This trend is also followed by the NSpM+Att2 results, while the remaining models, NSpM baseline, NSpM+Att1, and ConvS2S all show some deviations. These models achieve a higher performance in one or both measures on the sampled original questions compared to the full DBNQA$^*$-test split. Excepting the NSpM baseline, however, performance in both measures is less when predicting on the rewritten questions than the original questions. 
% In particular, these models achieve a higher performance in one or both measures on the sampled original questions compared to the full DBNQA$^*$-test split. However, performance in both measures is less when predicting on the rewritten questions than the original questions. The only exception is NSpM baseline, which has a very small benefit in EM from rewritten questions compared to original questions.  

We indicate statistical significance in Table~\ref{tab:results} on the performance of the rewritten questions sample compared to the original questions sample for each KGQA model. A single dagger ($\dag$) indicates that the $p$-value was below $\alpha=0.05$, while a double dagger ($\ddag$) indicates that the $p$-value was less than the Bonferroni-corrected threshold of $\frac{\alpha}{7}$ based on the seven comparisons made for each dependent variable.

% 09.02.2022 {
\if 0
%

% 
\fi
% } 09.02.2022
\section{Discussion}
\label{sec:discussion}

The present work addresses data quality and collects improved formulations of NL questions, to yield the IQN-KGQA test collection.  
We reflect on the data collection process, discuss possible uses of our test collection, and identify limitations.

\subsection{Data Collection} 

We have followed a similar procedure as the crowdsourced paraphrasing and cross-validation used in the construction  of several large-scale KGQA datasets described in Sect.~\ref{sec:related}. Unlike the reported crowdsourcing of prior datasets, we have involved the crowd workers in a consideration of language quality and question naturalness, by soliciting ratings on the five unnaturalness dimensions, immediately prior to paraphrasing an original NL question. The Likert ratings themselves provide a perspective into how crowd workers see NL questions that should be rewritten compared to those that should not or cannot be rewritten.

A majority of sampled NL questions were marked by some of the crowd workers as needing rewriting. Furthermore, during the second crowdsourced task, we see that the for most rewritten questions, the preferred version emerges quickly---in most cases with 3--5 votes. The resulting test collection has a majority (541 of 750) of its NL questions rewritten from their original form. This indicates that crowd workers have found room for improvement even after the initial paraphrasing and cross-validation undertaken in the original KGQA datasets' construction. This proportion of question rewrites also indicates that all three KGQA datasets can benefit in terms of question naturalness from extensive NL question rewriting. 

\subsection{Utilization}

The present work describes a process of improving NL questions for KGQA datasets. 
This shows the value of additional rounds of rewriting and quality control when creating NL questions via crowdsourcing. 
However, the reduced performance of KGQA models on the sample with rewritten NL questions also calls into question the overall approach of relying heavily on crowdsourcing for large scale KGQA dataset construction. 

We encourage other researchers to report their performance on our IQN-KGQA test collection as well as the test splits of the KGQA datasets on which they train their models. 
This will serve to keep the true KGQA performance in perspective. 
The reduced performance caused by rewritten NL questions illustrates that KGQA models are effectively overfitting on their datasets and do not generalize to natural question formulations. Our IQN-KGQA collection can be used to guard against this. 

Our crowdsourcing designs can be utilized in future large-scale KGQA dataset construction efforts.
The numerical ratings on the various dimensions of unnaturalness (in Task 1) may be used as quality control.
Our collection could also be utilized for automatic question rewriting using, e.g., for fine-tuning large language models, to generate question paraphrases to contribute to the pool of options that crowd workers can vote on (as in Task 2).

\subsection{Limitations}

We tried to simplify quality control of crowdsourcing by having some heuristics about what constituted a reasonable effort of rewrites, but these filters were perhaps not sufficient. 
There are examples where the crowdsourced rewriting and validation seem to fail to improve the NL question that is rewritten. For example, the original question ``Which past members of the labelle also sang somebody loves you baby (Blackstreet \& Ma song)?'' was voted down in favor of the rewritten question ``What song is Patti LaBelle famous for ?''

Our test collection is of small scale, yet has been relatively expensive to produce, on the order of US\$1000. 
Although crowdsourcing labor may be an economic way to scale up data annotation, there remains a question of how involved the manual quality control should be from the researchers' side.
\section{Conclusion}
\label{sec:conclusion}

We have investigated the dimensions of unnaturalness in the nominally natural language questions found in several modern large-scale knowledge graph question answering (KGQA) datasets. 
Specifically, we have developed a coding scheme to evaluate the naturalness of NL questions. 
We have also used crowdsourcing to rewrite such NL questions in KGQA datasets to be more genuinely natural. 
By combining language quality evaluation with NL question rewriting, we have attempted to prime crowd workers with attention towards language quality. 
From these rewritten NL questions, we have created the IQN-KGQA test collection with grounding in each of the three major knowledge graphs (KGs) addressed in previous KGQA research: DBpedia, Freebase, and Wikidata.   
This test collection can put KGQA performance in a more realistic perspective compared to testing KGQA systems on validation and test splits created with the exact same procedure as the training split. 
We have experimentally shown the impact of our test collection on the performance of KGQA models compared to performance on the corresponding sample of original NL questions and found that model performance deteriorated substantially when a more natural formulation of the same questions was provided.  This suggests that existing models do not generalize well to genuinely natural questions. 
The present work represents an initial effort to better understand ways to improve the naturalness of NL questions for KGQA and to ensure that KGQA performance is evaluated with genuinely natural questions. 

\bibliographystyle{ACM-Reference-Format}
\balance
\bibliography{sigir2022-datagen.bib}

%%% -*-BibTeX-*-
%%% Do NOT edit. File created by BibTeX with style
%%% ACM-Reference-Format-Journals [18-Jan-2012].

\begin{thebibliography}{33}

%%% ====================================================================
%%% NOTE TO THE USER: you can override these defaults by providing
%%% customized versions of any of these macros before the \bibliography
%%% command.  Each of them MUST provide its own final punctuation,
%%% except for \shownote{}, \showDOI{}, and \showURL{}.  The latter two
%%% do not use final punctuation, in order to avoid confusing it with
%%% the Web address.
%%%
%%% To suppress output of a particular field, define its macro to expand
%%% to an empty string, or better, \unskip, like this:
%%%
%%% \newcommand{\showDOI}[1]{\unskip}   % LaTeX syntax
%%%
%%% \def \showDOI #1{\unskip}           % plain TeX syntax
%%%
%%% ====================================================================

\ifx \showCODEN    \undefined \def \showCODEN     #1{\unskip}     \fi
\ifx \showDOI      \undefined \def \showDOI       #1{#1}\fi
\ifx \showISBNx    \undefined \def \showISBNx     #1{\unskip}     \fi
\ifx \showISBNxiii \undefined \def \showISBNxiii  #1{\unskip}     \fi
\ifx \showISSN     \undefined \def \showISSN      #1{\unskip}     \fi
\ifx \showLCCN     \undefined \def \showLCCN      #1{\unskip}     \fi
\ifx \shownote     \undefined \def \shownote      #1{#1}          \fi
\ifx \showarticletitle \undefined \def \showarticletitle #1{#1}   \fi
\ifx \showURL      \undefined \def \showURL       {\relax}        \fi
% The following commands are used for tagged output and should be
% invisible to TeX
\providecommand\bibfield[2]{#2}
\providecommand\bibinfo[2]{#2}
\providecommand\natexlab[1]{#1}
\providecommand\showeprint[2][]{arXiv:#2}

\bibitem[Arguello et~al\mbox{.}(2021)]%
        {Arguello:2021:CHIIR}
\bibfield{author}{\bibinfo{person}{Jaime Arguello}, \bibinfo{person}{Adam
  Ferguson}, \bibinfo{person}{Emery Fine}, \bibinfo{person}{Bhaskar Mitra},
  \bibinfo{person}{Hamed Zamani}, {and} \bibinfo{person}{Fernando Diaz}.}
  \bibinfo{year}{2021}\natexlab{}.
\newblock \showarticletitle{Tip of the Tongue Known-Item Retrieval: A Case
  Study in Movie Identification}. In \bibinfo{booktitle}{\emph{Proceedings of
  the 2021 Conference on Human Information Interaction and Retrieval}}
  \emph{(\bibinfo{series}{CHIIR '21})}. \bibinfo{pages}{5--14}.
\newblock


\bibitem[Bahdanau et~al\mbox{.}(2014)]%
        {Bahdanau:2014:arXiv}
\bibfield{author}{\bibinfo{person}{Dzmitry Bahdanau},
  \bibinfo{person}{Kyunghyun Cho}, {and} \bibinfo{person}{Yoshua Bengio}.}
  \bibinfo{year}{2014}\natexlab{}.
\newblock \bibinfo{title}{Neural Machine Translation by Jointly Learning to
  Align and Translate}.
\newblock
\newblock
\showeprint[arxiv]{1409.0473}~[cs.CL]


\bibitem[Bao et~al\mbox{.}(2016)]%
        {Bao:2016:COLING}
\bibfield{author}{\bibinfo{person}{Junwei Bao}, \bibinfo{person}{Nan Duan},
  \bibinfo{person}{Zhao Yan}, \bibinfo{person}{Ming Zhou}, {and}
  \bibinfo{person}{Tiejun Zhao}.} \bibinfo{year}{2016}\natexlab{}.
\newblock \showarticletitle{Constraint-Based Question Answering with Knowledge
  Graph}. In \bibinfo{booktitle}{\emph{Proceedings of {COLING} 2016, the 26th
  International Conference on Computational Linguistics: Technical Papers}}
  \emph{(\bibinfo{series}{COLING '16})}. \bibinfo{pages}{2503--2514}.
\newblock


\bibitem[Berant et~al\mbox{.}(2013)]%
        {Berant:2013:EMNLP}
\bibfield{author}{\bibinfo{person}{Jonathan Berant}, \bibinfo{person}{Andrew
  Chou}, \bibinfo{person}{Roy Frostig}, {and} \bibinfo{person}{Percy Liang}.}
  \bibinfo{year}{2013}\natexlab{}.
\newblock \showarticletitle{Semantic Parsing on {F}reebase from Question-Answer
  Pairs}. In \bibinfo{booktitle}{\emph{Proceedings of the 2013 Conference on
  Empirical Methods in Natural Language Processing}}
  \emph{(\bibinfo{series}{EMNLP '13})}. \bibinfo{pages}{1533--1544}.
\newblock


\bibitem[Bordes et~al\mbox{.}(2015)]%
        {Bordes:2015:arXiv}
\bibfield{author}{\bibinfo{person}{Antoine Bordes}, \bibinfo{person}{Nicolas
  Usunier}, \bibinfo{person}{Sumit Chopra}, {and} \bibinfo{person}{Jason
  Weston}.} \bibinfo{year}{2015}\natexlab{}.
\newblock \bibinfo{title}{Large-scale Simple Question Answering with Memory
  Networks}.
\newblock
\newblock
\showeprint[arxiv]{1506.02075}~[cs.LG]


\bibitem[Cai and Yates(2013)]%
        {Cai:2013:ACL}
\bibfield{author}{\bibinfo{person}{Qingqing Cai} {and}
  \bibinfo{person}{Alexander Yates}.} \bibinfo{year}{2013}\natexlab{}.
\newblock \showarticletitle{Large-scale Semantic Parsing via Schema Matching
  and Lexicon Extension}. In \bibinfo{booktitle}{\emph{Proceedings of the 51st
  Annual Meeting of the Association for Computational Linguistics (Volume 1:
  Long Papers)}} \emph{(\bibinfo{series}{ACL '13})}. \bibinfo{pages}{423--433}.
\newblock


\bibitem[Chakraborty et~al\mbox{.}(2021)]%
        {Chakraborty:2020:WIREsDMKD}
\bibfield{author}{\bibinfo{person}{Nilesh Chakraborty}, \bibinfo{person}{Denis
  Lukovnikov}, \bibinfo{person}{Gaurav Maheshwari}, \bibinfo{person}{Priyansh
  Trivedi}, \bibinfo{person}{Jens Lehmann}, {and} \bibinfo{person}{Asja
  Fischer}.} \bibinfo{year}{2021}\natexlab{}.
\newblock \showarticletitle{Introduction to Neural Network-based Question
  Answering over Knowledge Graphs}.
\newblock \bibinfo{journal}{\emph{WIREs Data Mining and Knowledge Discovery}}
  \bibinfo{volume}{11}, \bibinfo{number}{3} (\bibinfo{year}{2021}),
  \bibinfo{pages}{e1389}.
\newblock


\bibitem[Devlin et~al\mbox{.}(2019)]%
        {Devlin:2019:NAACL-HLT}
\bibfield{author}{\bibinfo{person}{Jacob Devlin}, \bibinfo{person}{Ming-Wei
  Chang}, \bibinfo{person}{Kenton Lee}, {and} \bibinfo{person}{Kristina
  Toutanova}.} \bibinfo{year}{2019}\natexlab{}.
\newblock \showarticletitle{{BERT}: Pre-training of Deep Bidirectional
  Transformers for Language Understanding}. In
  \bibinfo{booktitle}{\emph{Proceedings of the 2019 Conference of the North
  {A}merican Chapter of the Association for Computational Linguistics: Human
  Language Technologies, Volume 1 (Long and Short Papers)}}
  \emph{(\bibinfo{series}{NAACL-HLT '19})}. \bibinfo{pages}{4171--4186}.
\newblock


\bibitem[Dubey et~al\mbox{.}(2019)]%
        {Dubey:2019:ISWC}
\bibfield{author}{\bibinfo{person}{Mohnish Dubey}, \bibinfo{person}{Debayan
  Banerjee}, \bibinfo{person}{Abdelrahman Abdelkawi}, {and}
  \bibinfo{person}{Jens Lehmann}.} \bibinfo{year}{2019}\natexlab{}.
\newblock \showarticletitle{LC-QuAD 2.0: A Large Dataset for Complex Question
  Answering over Wikidata and DBpedia}. In
  \bibinfo{booktitle}{\emph{Proceedings of the 18th International Semantic Web
  Conference}} \emph{(\bibinfo{series}{ISWC '19})}. \bibinfo{pages}{69--78}.
\newblock


\bibitem[Gehring et~al\mbox{.}(2017)]%
        {Gehring:2017:ICML}
\bibfield{author}{\bibinfo{person}{Jonas Gehring}, \bibinfo{person}{Michael
  Auli}, \bibinfo{person}{David Grangier}, \bibinfo{person}{Denis Yarats},
  {and} \bibinfo{person}{Yann~N. Dauphin}.} \bibinfo{year}{2017}\natexlab{}.
\newblock \showarticletitle{Convolutional Sequence to Sequence Learning}. In
  \bibinfo{booktitle}{\emph{Proceedings of the 34th International Conference on
  Machine Learning - Volume 70}} \emph{(\bibinfo{series}{ICML'17})}.
  \bibinfo{pages}{1243--1252}.
\newblock


\bibitem[Gu et~al\mbox{.}(2021)]%
        {Gu:2021:WWW}
\bibfield{author}{\bibinfo{person}{Yu Gu}, \bibinfo{person}{Sue Kase},
  \bibinfo{person}{Michelle Vanni}, \bibinfo{person}{Brian Sadler},
  \bibinfo{person}{Percy Liang}, \bibinfo{person}{Xifeng Yan}, {and}
  \bibinfo{person}{Yu Su}.} \bibinfo{year}{2021}\natexlab{}.
\newblock \showarticletitle{Beyond I.I.D.: Three Levels of Generalization for
  Question Answering on Knowledge Bases}. In
  \bibinfo{booktitle}{\emph{Proceedings of the Web Conference 2021}}
  \emph{(\bibinfo{series}{WWW '21})}. \bibinfo{pages}{3477--3488}.
\newblock


\bibitem[Hartmann et~al\mbox{.}(2018)]%
        {Hartmann:2018:WEBBR-ws}
\bibfield{author}{\bibinfo{person}{Ann-Kathrin Hartmann},
  \bibinfo{person}{Edgard Marx}, {and} \bibinfo{person}{Tommaso Soru}.}
  \bibinfo{year}{2018}\natexlab{}.
\newblock \showarticletitle{Generating a Large Dataset for Neural Question
  Answering over the {DB}pedia Knowledge Base}. In
  \bibinfo{booktitle}{\emph{Workshop on Linked Data Management, co-located with
  the W3C WEBBR 2018}} \emph{(\bibinfo{series}{WEBBR '18})}.
\newblock


\bibitem[J\o{}rgensen and Bogers(2020)]%
        {Jorgensen:2020:FDG}
\bibfield{author}{\bibinfo{person}{Ida Kathrine~Hammeleff J\o{}rgensen} {and}
  \bibinfo{person}{Toine Bogers}.} \bibinfo{year}{2020}\natexlab{}.
\newblock \showarticletitle{``Kinda like The Sims... But with Ghosts?'': A
  Qualitative Analysis of Video Game Re-Finding Requests on Reddit}. In
  \bibinfo{booktitle}{\emph{International Conference on the Foundations of
  Digital Games}} \emph{(\bibinfo{series}{FDG '20})}.
\newblock


\bibitem[Keysers et~al\mbox{.}(2020)]%
        {Keysers:2020:ICLR}
\bibfield{author}{\bibinfo{person}{Daniel Keysers}, \bibinfo{person}{Nathanael
  Sch{\"a}rli}, \bibinfo{person}{Nathan Scales}, \bibinfo{person}{Hylke
  Buisman}, \bibinfo{person}{Daniel Furrer}, \bibinfo{person}{Sergii Kashubin},
  \bibinfo{person}{Nikola Momchev}, \bibinfo{person}{Danila Sinopalnikov},
  \bibinfo{person}{Lukasz Stafiniak}, \bibinfo{person}{Tibor Tihon},
  \bibinfo{person}{Dmitry Tsarkov}, \bibinfo{person}{Xiao Wang},
  \bibinfo{person}{Marc van Zee}, {and} \bibinfo{person}{Olivier Bousquet}.}
  \bibinfo{year}{2020}\natexlab{}.
\newblock \showarticletitle{Measuring Compositional Generalization: A
  Comprehensive Method on Realistic Data}. In
  \bibinfo{booktitle}{\emph{Proceedings of the 2020 International Conference on
  Learning Representations}} \emph{(\bibinfo{series}{ICLR '20})}.
\newblock


\bibitem[Lan et~al\mbox{.}(2021)]%
        {Lan:2021:IJCAI}
\bibfield{author}{\bibinfo{person}{Yunshi Lan}, \bibinfo{person}{Gaole He},
  \bibinfo{person}{Jinhao Jiang}, \bibinfo{person}{Jing Jiang},
  \bibinfo{person}{Wayne~Xin Zhao}, {and} \bibinfo{person}{Ji-Rong Wen}.}
  \bibinfo{year}{2021}\natexlab{}.
\newblock \showarticletitle{A Survey on Complex Knowledge Base Question
  Answering: Methods, Challenges and Solutions}. In
  \bibinfo{booktitle}{\emph{Proceedings of the Thirtieth International Joint
  Conference on Artificial Intelligence}} \emph{(\bibinfo{series}{IJCAI '21})}.
  \bibinfo{pages}{4483--4491}.
\newblock


\bibitem[Linjordet and Balog(2020)]%
        {Linjordet:2020:ICTIR}
\bibfield{author}{\bibinfo{person}{Trond Linjordet} {and}
  \bibinfo{person}{Krisztian Balog}.} \bibinfo{year}{2020}\natexlab{}.
\newblock \showarticletitle{Sanitizing Synthetic Training Data Generation for
  Question Answering over Knowledge Graphs}. In
  \bibinfo{booktitle}{\emph{Proceedings of the 2020 ACM SIGIR on International
  Conference on Theory of Information Retrieval}} \emph{(\bibinfo{series}{ICTIR
  '20})}. \bibinfo{pages}{121--128}.
\newblock


\bibitem[Lopez et~al\mbox{.}(2013)]%
        {Lopez:2013:WS}
\bibfield{author}{\bibinfo{person}{Vanessa Lopez}, \bibinfo{person}{Christina
  Unger}, \bibinfo{person}{Philipp Cimiano}, {and} \bibinfo{person}{Enrico
  Motta}.} \bibinfo{year}{2013}\natexlab{}.
\newblock \showarticletitle{Evaluating Question Answering over Linked Data}.
\newblock \bibinfo{journal}{\emph{Web Semant.: Science, Services and Agents on
  the World Wide Web}}  \bibinfo{volume}{21} (\bibinfo{year}{2013}),
  \bibinfo{pages}{3--13}.
\newblock


\bibitem[Luong et~al\mbox{.}(2015)]%
        {Luong:2015:arXiv}
\bibfield{author}{\bibinfo{person}{Minh-Thang Luong}, \bibinfo{person}{Hieu
  Pham}, {and} \bibinfo{person}{Christopher~D Manning}.}
  \bibinfo{year}{2015}\natexlab{}.
\newblock \bibinfo{title}{Effective Approaches to Attention-based Neural
  Machine Translation}.
\newblock
\newblock
\showeprint[arxiv]{1508.04025}~[cs.CL]


\bibitem[Roy and Anand(2021)]%
        {Roy:2021:SLcICRS}
\bibfield{author}{\bibinfo{person}{Rishiraj~Saha Roy} {and}
  \bibinfo{person}{Avishek Anand}.} \bibinfo{year}{2021}\natexlab{}.
\newblock \showarticletitle{Question Answering for the Curated Web: Tasks and
  Methods in QA over Knowledge Bases and Text Collections}.
\newblock \bibinfo{journal}{\emph{Synthesis Lectures on Inf. Concepts, Retr.,
  and Services}} \bibinfo{volume}{13}, \bibinfo{number}{4}
  (\bibinfo{year}{2021}), \bibinfo{pages}{1--194}.
\newblock


\bibitem[Sennrich et~al\mbox{.}(2015)]%
        {Sennrich:2015:arXiv}
\bibfield{author}{\bibinfo{person}{Rico Sennrich}, \bibinfo{person}{Barry
  Haddow}, {and} \bibinfo{person}{Alexandra Birch}.}
  \bibinfo{year}{2015}\natexlab{}.
\newblock \bibinfo{title}{Neural Machine Translation of Rare Words with Subword
  Units}.
\newblock
\newblock
\showeprint[arxiv]{1508.07909}~[cs.CL]


\bibitem[Shi et~al\mbox{.}(2020)]%
        {Shi:2020:arXiv}
\bibfield{author}{\bibinfo{person}{Jiaxin Shi}, \bibinfo{person}{Shulin Cao},
  \bibinfo{person}{Liangming Pan}, \bibinfo{person}{Yutong Xiang},
  \bibinfo{person}{Lei Hou}, \bibinfo{person}{Juanzi Li},
  \bibinfo{person}{Hanwang Zhang}, {and} \bibinfo{person}{Bin He}.}
  \bibinfo{year}{2020}\natexlab{}.
\newblock \bibinfo{title}{{KQA} Pro: {A} Large Diagnostic Dataset for Complex
  Question Answering over Knowledge Base}.
\newblock
\newblock
\showeprint[arxiv]{2007.03875}~[cs.CL]


\bibitem[Soru et~al\mbox{.}(2017)]%
        {Soru:2017:arXiv}
\bibfield{author}{\bibinfo{person}{Tommaso Soru}, \bibinfo{person}{Edgard
  Marx}, \bibinfo{person}{Diego Moussallem}, \bibinfo{person}{Gustavo Publio},
  \bibinfo{person}{Andr{\'e} Valdestilhas}, \bibinfo{person}{Diego Esteves},
  {and} \bibinfo{person}{Ciro~Baron Neto}.} \bibinfo{year}{2017}\natexlab{}.
\newblock \showarticletitle{SPARQL as a Foreign Language}.
\newblock  (\bibinfo{year}{2017}).
\newblock
\showeprint[arxiv]{1708.07624}~[cs.CL]


\bibitem[Soru et~al\mbox{.}(2018)]%
        {Soru:2018:ICML-ws}
\bibfield{author}{\bibinfo{person}{Tommaso Soru}, \bibinfo{person}{Edgard
  Marx}, \bibinfo{person}{Andr\'e Valdestilhas}, \bibinfo{person}{Diego
  Esteves}, \bibinfo{person}{Diego Moussallem}, {and} \bibinfo{person}{Gustavo
  Publio}.} \bibinfo{year}{2018}\natexlab{}.
\newblock \showarticletitle{Neural Machine Translation for Query Construction
  and Composition}. In \bibinfo{booktitle}{\emph{ICML Workshop on Neural
  Abstract Machines \& Program Induction (NAMPI v2)}}
  \emph{(\bibinfo{series}{ICML '18})}.
\newblock


\bibitem[Su et~al\mbox{.}(2016)]%
        {Su:2016:EMNLP}
\bibfield{author}{\bibinfo{person}{Yu Su}, \bibinfo{person}{Huan Sun},
  \bibinfo{person}{Brian Sadler}, \bibinfo{person}{Mudhakar Srivatsa},
  \bibinfo{person}{Izzeddin G{\"u}r}, \bibinfo{person}{Zenghui Yan}, {and}
  \bibinfo{person}{Xifeng Yan}.} \bibinfo{year}{2016}\natexlab{}.
\newblock \showarticletitle{On Generating Characteristic-rich Question Sets for
  {QA} Evaluation}. In \bibinfo{booktitle}{\emph{Proceedings of the 2016
  Conference on Empirical Methods in Natural Language Processing}}
  \emph{(\bibinfo{series}{EMNLP '16})}. \bibinfo{pages}{562--572}.
\newblock


\bibitem[Talmor and Berant(2018)]%
        {Talmor:2018:NAACL}
\bibfield{author}{\bibinfo{person}{Alon Talmor} {and} \bibinfo{person}{Jonathan
  Berant}.} \bibinfo{year}{2018}\natexlab{}.
\newblock \showarticletitle{The Web as a Knowledge-Base for Answering Complex
  Questions}. In \bibinfo{booktitle}{\emph{Proceedings of the 2018 Conference
  of the North {A}merican Chapter of the Association for Computational
  Linguistics: Human Language Technologies, Volume 1 (Long Papers)}}
  \emph{(\bibinfo{series}{NAACL '18})}. \bibinfo{pages}{641--651}.
\newblock


\bibitem[Trivedi et~al\mbox{.}(2017)]%
        {Trivedi:2017:ISWC}
\bibfield{author}{\bibinfo{person}{Priyansh Trivedi}, \bibinfo{person}{Gaurav
  Maheshwari}, \bibinfo{person}{Mohnish Dubey}, {and} \bibinfo{person}{Jens
  Lehmann}.} \bibinfo{year}{2017}\natexlab{}.
\newblock \showarticletitle{{LC-QuAD}: A Corpus for Complex Question Answering
  over Knowledge Graphs}. In \bibinfo{booktitle}{\emph{Proceedings of the 16th
  International Semantic Web Conference (ISWC)}} \emph{(\bibinfo{series}{ISWC
  '17})}. \bibinfo{pages}{210--218}.
\newblock


\bibitem[Usbeck et~al\mbox{.}(2017)]%
        {Usbeck:2017:ESWC}
\bibfield{author}{\bibinfo{person}{Ricardo Usbeck},
  \bibinfo{person}{Axel-Cyrille Ngonga~Ngomo}, \bibinfo{person}{Bastian
  Haarmann}, \bibinfo{person}{Anastasia Krithara}, \bibinfo{person}{Michael
  R{\"o}der}, {and} \bibinfo{person}{Giulio Napolitano}.}
  \bibinfo{year}{2017}\natexlab{}.
\newblock \showarticletitle{7th Open Challenge on Question Answering over
  Linked Data (QALD-7)}. In \bibinfo{booktitle}{\emph{Proceedings of Semantic
  Web Challenges - 4th SemWebEval Challenge at {ESWC} 2017}}
  \emph{(\bibinfo{series}{ESWC '17}, Vol.~\bibinfo{volume}{769})}.
  \bibinfo{pages}{59--69}.
\newblock


\bibitem[Usbeck et~al\mbox{.}(2019)]%
        {Usbeck:2019:SW}
\bibfield{author}{\bibinfo{person}{Ricardo Usbeck}, \bibinfo{person}{Michael
  R{\"o}der}, \bibinfo{person}{Michael Hoffmann}, \bibinfo{person}{Felix
  Conrads}, \bibinfo{person}{Jonathan Huthmann}, \bibinfo{person}{Axel-Cyrille
  Ngonga-Ngomo}, \bibinfo{person}{Christian Demmler}, {and}
  \bibinfo{person}{Christina Unger}.} \bibinfo{year}{2019}\natexlab{}.
\newblock \bibinfo{journal}{\emph{Semantic Web}} \bibinfo{volume}{10},
  \bibinfo{number}{2} (\bibinfo{year}{2019}), \bibinfo{pages}{293--304}.
\newblock


\bibitem[Vaswani et~al\mbox{.}(2017)]%
        {Vaswani:2017:NeurIPS}
\bibfield{author}{\bibinfo{person}{Ashish Vaswani}, \bibinfo{person}{Noam
  Shazeer}, \bibinfo{person}{Niki Parmar}, \bibinfo{person}{Jakob Uszkoreit},
  \bibinfo{person}{Llion Jones}, \bibinfo{person}{Aidan~N. Gomez},
  \bibinfo{person}{\L{}ukasz Kaiser}, {and} \bibinfo{person}{Illia
  Polosukhin}.} \bibinfo{year}{2017}\natexlab{}.
\newblock \showarticletitle{Attention is All You Need}. In
  \bibinfo{booktitle}{\emph{Proceedings of the 31st International Conference on
  Neural Information Processing Systems}} \emph{(\bibinfo{series}{NeurIPS
  '17})}. \bibinfo{pages}{6000--6010}.
\newblock


\bibitem[Wang et~al\mbox{.}(2015)]%
        {Wang:2015:ACL-IJNLP}
\bibfield{author}{\bibinfo{person}{Yushi Wang}, \bibinfo{person}{Jonathan
  Berant}, {and} \bibinfo{person}{Percy Liang}.}
  \bibinfo{year}{2015}\natexlab{}.
\newblock \showarticletitle{Building a Semantic Parser Overnight}. In
  \bibinfo{booktitle}{\emph{Proceedings of the 53rd Annual Meeting of the
  Association for Computational Linguistics and the 7th International Joint
  Conference on Natural Language Processing}} \emph{(\bibinfo{series}{ACL-IJNLP
  '15})}. \bibinfo{pages}{1332--1342}.
\newblock


\bibitem[Wu et~al\mbox{.}(2019)]%
        {Wu:2019:CCKS}
\bibfield{author}{\bibinfo{person}{Peiyun Wu}, \bibinfo{person}{Xiaowang
  Zhang}, {and} \bibinfo{person}{Zhiyong Feng}.}
  \bibinfo{year}{2019}\natexlab{}.
\newblock \showarticletitle{A Survey of Question Answering over Knowledge
  Base}. In \bibinfo{booktitle}{\emph{Knowledge Graph and Semantic Computing:
  Knowledge Computing and Language Understanding}}. \bibinfo{pages}{86--97}.
\newblock


\bibitem[Yih et~al\mbox{.}(2016)]%
        {Yih:2016:ACL}
\bibfield{author}{\bibinfo{person}{Wen-tau Yih}, \bibinfo{person}{Matthew
  Richardson}, \bibinfo{person}{Chris Meek}, \bibinfo{person}{Ming-Wei Chang},
  {and} \bibinfo{person}{Jina Suh}.} \bibinfo{year}{2016}\natexlab{}.
\newblock \showarticletitle{The Value of Semantic Parse Labeling for Knowledge
  Base Question Answering}. In \bibinfo{booktitle}{\emph{Proceedings of the
  54th Annual Meeting of the Association for Computational Linguistics (Volume
  2: Short Papers)}} \emph{(\bibinfo{series}{ACL '16})}.
  \bibinfo{pages}{201--206}.
\newblock


\bibitem[Yin et~al\mbox{.}(2021)]%
        {Yin:2021:FGCS}
\bibfield{author}{\bibinfo{person}{Xiaoyu Yin}, \bibinfo{person}{Dagmar
  Gromann}, {and} \bibinfo{person}{Sebastian Rudolph}.}
  \bibinfo{year}{2021}\natexlab{}.
\newblock \showarticletitle{Neural Machine Translating from Natural Language to
  {SPARQL}}.
\newblock \bibinfo{journal}{\emph{Future Generation Computer Systems}}
  \bibinfo{volume}{117} (\bibinfo{year}{2021}), \bibinfo{pages}{510--519}.
\newblock


\end{thebibliography}

\end{document}